\pdfoutput=1
\documentclass[twocolumn]{aastex701}

\usepackage{amsmath}
\usepackage{bm}

\newcommand{\Mdot}{{\rm M}_{\odot}}

\newcommand{\angstrom}{\text{\AA}}

\graphicspath{{./}{images/}}

\begin{document}

\title{Reconstructing the Three-Dimensional Stellar Mass Profiles of Massive Galaxy Clusters}


\author[0009-0009-4334-5598]{Jong-In Park}
\affiliation{Astronomy Program, Department of Physics and Astronomy, Seoul National University, 1 Gwanak-ro, Gwanak-gu, Seoul 08826, Republic of Korea}
\email{jongin.park@snu.ac.kr}

\author[0000-0002-9254-144X]{Jubee Sohn}
\affiliation{Astronomy Program, Department of Physics and Astronomy, Seoul National University, 1 Gwanak-ro, Gwanak-gu, Seoul 08826, Republic of Korea}
\affiliation{SNU Astronomy Research Center, Seoul National University, Seoul 08826, Republic of Korea}
\email{jubee.sohn@snu.ac.kr}

\author[0000-0002-5819-6566]{I-Non Chiu}
\affiliation{Department of Physics, National Cheng Kung University, No.1, University Road, Tainan City 70101, Taiwan}
\email{inchiu@phys.ncku.edu.tw}

\begin{abstract}
We investigate the three-dimensional (3D) stellar mass profiles of galaxy clusters using simulations and observations. We develop the Projection-to-Enclosed Estimator by Layers (PEEL) model, which geometrically reconstructs the 3D stellar mass profile from the projected stellar mass profile under the assumption of spherical symmetry. Starting from the outermost annulus, the PEEL model subtracts the contributions of the outer shells from the projected mass in each inner annulus. Using 205 halos from the TNG-Cluster simulation, we show that the reconstructed profile (hereafter the PEEL profile) recovers the true 3D stellar mass profile without systematic bias. We also examine potential systematic effects, including the use of subhalos versus stellar particles, the choice of cluster center, and the 3D shape of the halo. We then apply the PEEL model to a stacked sample of nine massive clusters from the MAssive Cluster survey with Hectospec (MACH). The observed PEEL profile is much flatter than the simulated 3D profile, indicating that the simulated clusters have a more centrally concentrated stellar mass distribution, mainly because of an excess of extremely massive galaxies in their cores. Combining the PEEL profile with the total mass profile from the caustic technique, we derive the stellar mass fraction profile of the observed clusters. The stellar mass fraction remains nearly constant at $0.009 \pm 0.002$ within $R_{200}$, consistent with previous observations and with the simulated clusters. The PEEL model provides a practical tool for measuring the 3D stellar mass profiles of large cluster samples from future wide-field spectroscopic surveys. 
\end{abstract}


\section{Introduction}

The mass composition of galaxy clusters provides an important test of cosmological structure formation. In the standard $\Lambda$CDM model, dark matter (DM) halos form first, and their gravitational potential wells accrete baryons that subsequently cool to form stars and galaxies. Because clusters are the most massive collapsed structures in the universe, their baryon content records how baryons are accreted, retained, and processed during the assembly of massive systems. In particular, the partition of baryons between hot gas and stars reflects the efficiency of star formation across different mass scales (e.g., \citealp{Lin2003, Gonzalez2007, Lagana2008, Gonzalez2013}).

The baryon fraction of galaxy groups and clusters has been measured with multi-wavelength observations. The total mass of a system is often derived from X-ray observations \citep{Lin2003, Lagana2008, Gonzalez2013}, the Sunyaev-Zel'dovich (SZ) effect \citep{Chiu2016a, Chiu2018b}, or gravitational lensing \citep{Mulroy2019, Akino2022, Kim2026}. X-ray observations also provide the mass of the hot intracluster gas (e.g., \citealp{Lin2003, Gonzalez2013, Chiu2016a, Chiu2018b, Akino2022}). Estimating the total stellar mass is less straightforward, because identifying cluster members and estimating their stellar masses require extensive spectroscopy. The total stellar mass is therefore often estimated by integrating the luminosity or stellar mass function of the cluster galaxies. Despite this challenge, the stellar mass fraction provides unique information on the integrated efficiency of star formation over the assembly history of a system (e.g., \citealp{Lin2003, Leauthaud2012, Kravtsov2018}).

The stellar mass fraction of galaxy clusters, $f_{\star}$, is only $\sim 1$--$2\%$ of the total mass. Given the cosmic baryon fraction of $\Omega_\mathrm{b}/\Omega_\mathrm{m} \approx 0.16$ \citep{Ade2016}, only $\sim 10$--$20\%$ of the baryons in clusters reside in stars \citep{Lin2003, Gonzalez2013}. Comparing the gas and stellar mass fractions therefore constrains the efficiency of star formation and the recycling of gas in galaxy systems.

Previous measurements of $f_{\star}$ usually provide an integrated value within a fixed aperture, such as $R_{500}$. \citet{Lin2003} measured $f_{\star}$ of clusters using X-ray data and the $K$-band luminosity function, and found that $f_{\star}$ decreases from $2.2\%$ to $1.2\%$ as the cluster mass increases from $M_{500} = 10^{14}~\Mdot$ to $10^{15}~\Mdot$. \citet{Giodini2009} extended this measurement to the group scale and found $f_{\star} \sim 6\%$ for groups with $M_{500} \sim 2 \times 10^{13}~\Mdot$. More recently, \citet{Chiu2016a} and \citet{Chiu2018b} measured $f_{\star}$ for SZ-selected clusters over a wider redshift range of $0.2$--$1.3$. They found typical values of $f_{\star} \sim 0.83$--$1.1\%$ for clusters with $M_{500} = (4.8$--$6.0) \times 10^{14}~\Mdot$, with little dependence on redshift.

The radial profile of $f_{\star}$ provides additional information on the assembly of clusters. In cosmological simulations, $f_{\star}$ depends on cluster-centric radius, with a high concentration of stellar mass in the cluster center, including the brightest cluster galaxy (BCG; e.g., \citealp{Planelles2013, Sampaio-Santos2021}). However, these predictions have not been tested against observations, because only a few $f_{\star}$ profiles have been measured. For example, \citet{Annunziatella2014} derived the three-dimensional (3D) stellar mass density profile of MACS~J1206.2$-$0847 at $z = 0.44$ using 1363 members with stellar masses above $10^{9.5}~\Mdot$, but only about one third of these members were spectroscopically confirmed. Clusters with dense spectroscopic coverage therefore offer an ideal testbed for calibrating the stellar mass fraction profiles in simulations and for understanding the assembly history of galaxy clusters.

Here, we investigate the 3D stellar mass profiles of galaxy clusters. Observations provide only the projected two-dimensional (2D) stellar mass profile, whereas the physically relevant quantity is the 3D profile. We therefore develop the Projection-to-Enclosed Estimator by Layers (PEEL) model, which reconstructs the 3D stellar mass profile from the 2D projected profile under the assumption of an isolated, spherically symmetric stellar mass distribution. Similar deprojection techniques have been used to study the number density profiles of globular cluster systems around elliptical galaxies \citep{McLaughlin1999} and the X-ray surface brightness profiles of galaxy clusters and groups \citep{Fabian1981, Kriss1983}. We first test the PEEL model with the TNG-Cluster simulation \citep{Nelson2024}, which provides the 3D positions of cluster galaxies and allows us to evaluate potential systematic effects in the model. We then apply the model to clusters with extremely dense spectroscopy from the MAssive Cluster survey with Hectospec (MACH; \citealp{Park2026b}). Finally, we derive the stellar mass fraction profile of the observed clusters and compare it with the simulations.

We describe the PEEL model in Section~\ref{sec:peel_model}. Section~\ref{sec:data} describes the observational data and the simulations used in this work. In Section~\ref{sec:simulation_test}, we test the PEEL model with the TNG-Cluster halos and discuss potential sources of systematic uncertainty. Section~\ref{sec:comparison} presents the 3D stellar mass profiles of the observed clusters and compares them with the simulated profiles. We summarize our results in Section~\ref{sec:conclusion}. Throughout this paper, we adopt a flat $\Lambda$CDM cosmology with $H_0 = 70\,\mathrm{km\,s^{-1}\,Mpc^{-1}}$, $\Omega_\Lambda = 0.7$, and $\Omega_\mathrm{m} = 0.3$. We also quote the $1\sigma $ uncertainties and error bars throughout the paper unless stated otherwise. 

\section{The PEEL Model Framework} \label{sec:peel_model}

We reconstruct the 3D stellar mass profile of a galaxy cluster from the observable 2D projected stellar mass profile. We first compute the total stellar mass in a series of projected annuli. Each projected annulus contains stellar mass from the 3D shell at the same radius and from all outer shells that intersect it along the line of sight. Starting from the outermost annulus, we therefore derive the stellar mass in each 3D shell by subtracting the contributions of the outer shells, weighted by the volume they share with the corresponding projected annulus. Because this procedure removes the mass contributions layer by layer from the outside inward, we name our model the Projection-to-Enclosed Estimator by Layers (PEEL) model. The PEEL model relies on two key assumptions: (1) the stellar mass distribution of the cluster is spherically symmetric, and (2) the mass density beyond the outermost shell is negligible, so that the surrounding field makes no contribution to the projected mass profile.

Similar approaches have been used to reconstruct 3D distributions from projected distributions for various astronomical systems. \citet{Fabian1981} and \citet{Kriss1983} applied this technique to the X-ray surface brightness profiles of galaxy clusters and groups. \citet{McLaughlin1999} applied a similar technique to the number density profiles of globular clusters around elliptical galaxies.

\begin{figure}[tp]
    \centering
    \includegraphics[width=\columnwidth]{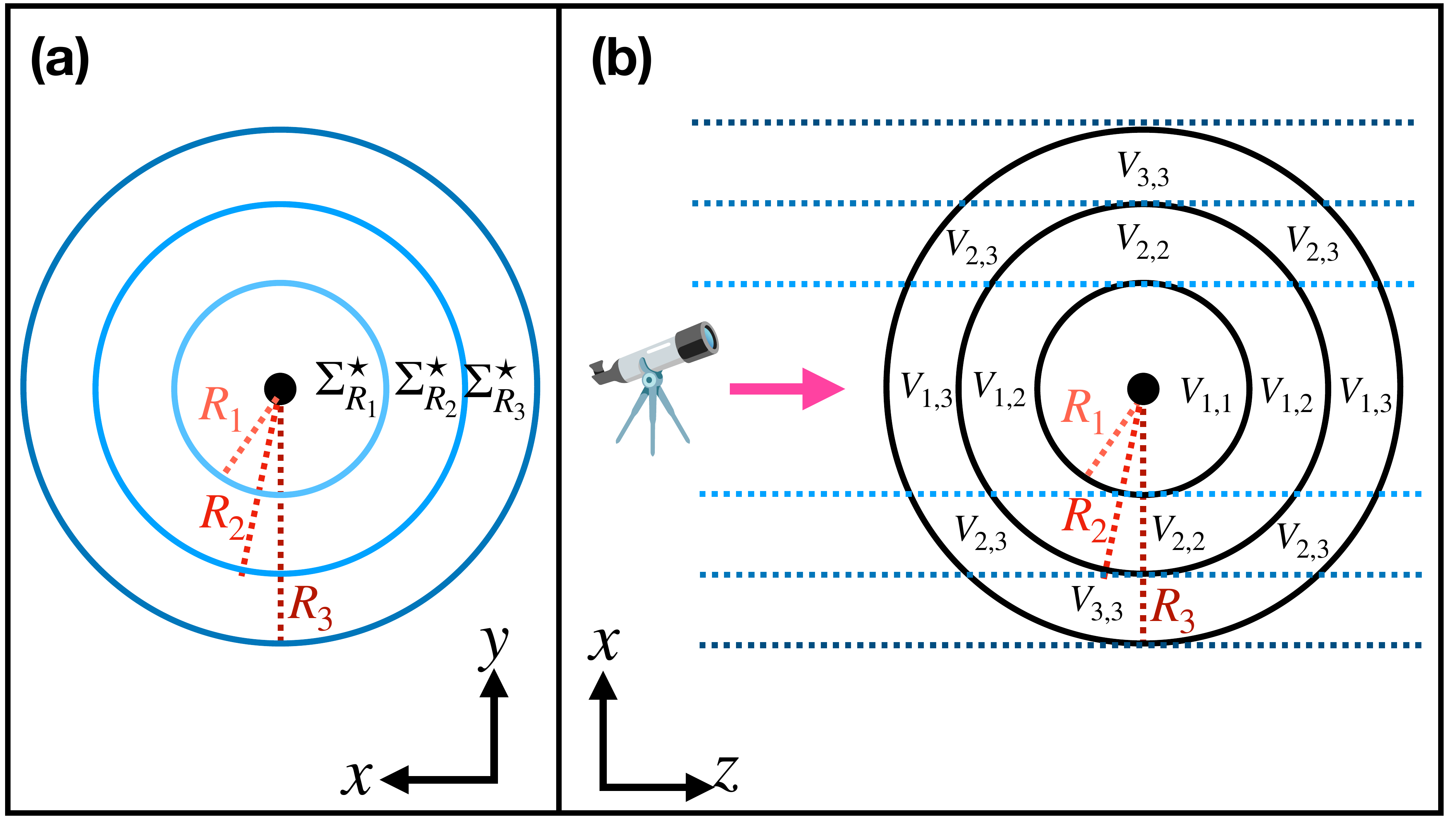}
        \caption{Schematic diagram of the PEEL model. (a)~Projected (plane-of-sky) view. The cluster is divided into concentric annuli with edge radii $R_\alpha$ ($R_0 = 0$); $\Sigma^\star_{R_\alpha}$ is the stellar mass summed within the annulus $R_{\alpha-1} \leq r_\mathrm{2D} < R_\alpha$. (b)~Bird's-eye (3D) view. The same radii define concentric spherical shells. The telescope and the arrow mark the line of sight, and the dotted horizontal lines mark the cylinders defined by the projected annuli. $V_{\alpha,\beta}$ denotes the volume of the portion of shell $\beta$ enclosed by the cylinder of annulus $\alpha$.}
        \label{fig:PEEL_diagram}
\end{figure}

Figure~\ref{fig:PEEL_diagram} shows a schematic view of the PEEL model. Figure~\ref{fig:PEEL_diagram}(a) displays a cluster on a projected plane. Observationally, the projected mass profile is derived by computing the sum of mass within the concentric annuli. The PEEL model first constructs the projected mass profile in the same way, by summing the stellar mass within each concentric annulus. We denote the edge radii of the projected annuli by $R_\alpha$, with $R_0 = 0$. $\Sigma^\star_{R_\alpha}$ indicates the summation of stellar mass within the projected annulus with $R_{\alpha-1} \leq r_\mathrm{2D} < R_{\alpha}$.

Figure~\ref{fig:PEEL_diagram}(b) shows the same cluster in the bird's-eye view. Each annulus in panel (b) indicates the 3D spherical shell. Comparing panels (a) and (b), the multiple segments of the 3D spherical shells contribute their enclosed mass to the projected mass profile. We define the overlapping volume $V_{\alpha,\beta}$ as the volume of the intersection between the spherical shell $\beta$ (with $R_{\beta-1} \leq r_\mathrm{3D} < R_\beta$) and the line-of-sight cylindrical annulus $\alpha$ (with $R_{\alpha-1} \leq r_\mathrm{2D} < R_\alpha$). The first index denotes the projected annulus and the second index denotes the spherical shell. For example, at the cluster core region, $\Sigma^\star_{R_1}$ is the sum of the mass enclosed in $V_{1,1}, V_{1,2}, \ldots, V_{1,N}$. Based on these relations, we reconstruct the 3D spherical mass profile from the projected mass profile. 

$\mathcal{M}_\star^{R_\beta}$ denotes the stellar mass in the spherical shell $\beta$ with $R_{\beta-1} \leq r_\mathrm{3D} < R_{\beta}$. The volume of the spherical shell $\beta$ is $\sum_{k=1}^N V_{k,\beta}$. We then define the stellar mass density for shell $\beta$ as $\rho_{\beta} = \mathcal{M}_{\star}^{R_\beta} / \sum_{k=1}^N V_{k,\beta}$, assuming that the density is constant within each shell. 

We express the observed stellar mass $\Sigma^{\star}_{R_\alpha}$ as the sum of the contributions from all 3D shells along the line of sight:
\begin{equation}
\label{eq:projected_mass_formulation}
\Sigma^{\star}_{R_\alpha} = \sum_{k=1}^{N} \rho_{k} V_{\alpha,k}.
\end{equation}
We note that the total stellar mass in the 3D spherical shell $\beta$ is the sum of contributions from all line-of-sight cylinders:
\begin{equation}
\label{eq:3d_mass_formulation}
\mathcal{M}_{\star}^{R_\beta} = \rho_{\beta} \sum_{k=1}^{N} V_{k,\beta}.
\end{equation}

We relate the projected mass (Equation~\ref{eq:projected_mass_formulation}) and the 3D mass (Equation~\ref{eq:3d_mass_formulation}) based on the density $\rho$. We treat the geometric factors $V_{\alpha,\beta}$ as coefficients in a linear system. We solve the linear system to eliminate $\rho$. We derive the stellar mass profile $\mathcal{M}_{\star}^{R_i}$ directly as a function of the observed 2D projected stellar mass profile $\Sigma^{\star}_{R_j}$ and the geometric projection matrix.

Combining Equations~\ref{eq:projected_mass_formulation} and~\ref{eq:3d_mass_formulation}, the PEEL stellar mass profile (hereafter the PEEL profile) $\mathcal{M}_{\star}^{R_i}$ takes the form:

\begin{equation}
\mathcal{M}_\star^{R_i} = \left( \sum_{j=i}^N \left[ V^{-1} \right]_{i,j} \Sigma^\star_{R_j} \right) \left( \sum_{k=1}^i V_{k,i} \right)
\end{equation}

We compute elements of the volume matrix $V_{i,j}$ from the iterative subtraction of the relevant spherical volumes:
\begin{align}
    V_{i,j} &= \left[ D_g(R_{i-1}, R_{j}) - D_g(R_{i}, R_{j}) \right] \nonumber \\
    &\quad - \left[ D_g(R_{i-1}, R_{j-1}) - D_g(R_{i}, R_{j-1}) \right], \nonumber \\
    &\quad R_0 = 0 \label{eq:volume_matrix} \\
    D_g(R_a, R_b) &= \begin{cases}
        \frac{4}{3}\pi \left(R_b^2 - R_a^2\right)^{\frac{3}{2}} & (R_a < R_b) \\
        0 & (R_a \geq R_b)
    \end{cases}
\end{align}
We present the geometric derivation of Equation~\ref{eq:volume_matrix} in Appendix~\ref{app:volume_matrix}.

The PEEL profile can be rewritten as Equation~\ref{eq:peel_matrix}:
\onecolumngrid
\begin{equation}
\label{eq:peel_matrix}
\renewcommand{\arraystretch}{1.4}
\begin{bmatrix}
\mathcal{M}_\star^{R_1} \\
\mathcal{M}_\star^{R_2} \\
\vdots \\
\mathcal{M}_\star^{R_{N-1}} \\
\mathcal{M}_\star^{R_N}
\end{bmatrix}
=
\left(
\begin{bmatrix}
V_{1,1} & V_{1,2} & \cdots & V_{1,N-1} & V_{1,N} \\
0 & V_{2,2} & \cdots & V_{2,N-1} & V_{2,N} \\
\vdots & \vdots & \ddots & \vdots & \vdots \\
0 & 0 & \cdots & V_{N-1,N-1} & V_{N-1,N} \\
0 & 0 & \cdots & 0 & V_{N,N}
\end{bmatrix}^{-1}
\begin{bmatrix}
\Sigma^\star_{R_1} \\
\Sigma^\star_{R_2} \\
\vdots \\
\Sigma^\star_{R_{N-1}} \\
\Sigma^\star_{R_N}
\end{bmatrix}
\right)
\odot
\begin{bmatrix}
V_{1,1} \\
V_{1,2} + V_{2,2} \\
\vdots \\
V_{1,N-1} + V_{2,N-1} + \cdots + V_{N-1,N-1} \\
V_{1,N} + V_{2,N} + \cdots + V_{N,N}
\end{bmatrix}
\end{equation}
\twocolumngrid
\noindent where the symbol $\odot$ denotes the element-wise product of two column vectors: the $i$-th component of $\mathbf{a} \odot \mathbf{b}$ is $a_i b_i$.

\section{Data and Simulations} \label{sec:data}

We describe the simulated and observational data sets used in this study. In Section~\ref{sec:TNG_Cluster}, we introduce the TNG-Cluster simulation, where we obtain the cluster sample for testing the PEEL model. In Section~\ref{sec:MACH_survey}, we describe the MACH survey, which provides the observational data. We summarize the key properties of each data set, including the cluster sample and the derived stellar mass measurements.

\subsection{The TNG-Cluster Simulation} \label{sec:TNG_Cluster}

TNG-Cluster \citep{Nelson2024} is an extended simulation of the IllustrisTNG project (hereafter TNG; \citealp{Nelson2018, Pillepich2018a, Marinacci2018, Springel2018, Naiman2018}). TNG-Cluster uses the moving-mesh code AREPO \citep{Springel2010} and adopts the same galaxy formation physics as TNG. The physics model includes thermal and kinetic active galactic nucleus (AGN) feedback modes \citep{Weinberger2017}, radiative cooling, star formation, galactic winds, and metal enrichment \citep{Pillepich2018b}. The TNG model reproduces the observed properties of galaxy clusters across a wide range of contexts (e.g., \citealp{Vogelsberger2018, Nelson2024, Park2026b}) and generates realistic stellar and gas content for both BCGs (e.g., \citealp{Montenegro-Taborda2023}) and satellite galaxies (e.g., \citealp{Stevens2019, Donnari2021, Park2026b, Park2026a}).

TNG-Cluster first identifies cluster halos as friends-of-friends (FoF) groups in a gravity-only parent simulation of a $\sim 1~{\rm Gpc}^{3}$ volume. Then, a subset of these clusters, selected by halo mass, is re-simulated with the high-resolution zoom-in technique. The zoom-in simulation has nearly the same mass resolution as the IllustrisTNG300-1 simulation, with the DM particle mass $m_\mathrm{DM} = 6.1 \times 10^{7}\,\Mdot$, and the baryonic particle mass $m_\mathrm{baryon} = 1.2 \times 10^7\,\Mdot$. The gravitational softening length of TNG-Cluster zoom-in simulation is $\sim 1.5$~kpc at $z=0$ for collisionless (i.e., DM and star) particles; gas cells have an adaptive softening with a minimum of 370 comoving pc. We adopt the position of the central subhalo of the target cluster halo as the cluster center for computing cluster-centric distances. 

We use TNG-Cluster halos with properties similar to those of the observed MACH clusters (Section~\ref{sec:MACH_survey}). To match the redshift range of the observed sample, we use snapshot 93 ($z \approx 0.07$) of the TNG-Cluster. In snapshot 93, there are 251 cluster halos with mass $\log ({M_{200}/\Mdot}) \gtrsim 14.7$, a mass range similar to that of the MACH cluster sample. However, not all of them have high-resolution zoom-in simulations. Additionally, there are a few clusters containing low-resolution DM particles ($0.29$--$30\%$) among the zoom-in clusters. We finally select 205 primary zoom-in target clusters (i.e., \texttt{GroupPrimaryZoomTarget} $= 1$) with $10^{14.7}\,\Mdot \leq M_{200} \leq 10^{15.3}\,\Mdot$. 

\subsection{The MACH Survey} \label{sec:MACH_survey}

The MACH survey \citep{Sohn2020, Park2026b} is a dense spectroscopic survey of the nine most massive galaxy clusters (A7, A1767, A1831, A2029, A2034, A2065, A2244, A2245, and A2255) in the local universe at $0.07 < z < 0.11$. MACH acquires over 4500 spectra for each target cluster. The MACH clusters have dynamical masses ranging from $5.5 \times 10^{14}\,\Mdot < M_{200} < 1.0 \times 10^{15}\,\Mdot$, with a mean mass of $7.4 \times 10^{14}\,\Mdot$. The cluster mass $M_{200}$ is the mass enclosed within the radius where the mean interior density is 200 times the critical density of the universe, derived using the caustic technique \citep{Diaferio1997, Diaferio1999, Serra2013}. \citet{Park2026b} describe the MACH survey in detail.

The MACH survey uses photometric catalogs from the Sloan Digital Sky Survey Data Release 18 (SDSS DR18; \citealp{Almeida2023}). The MACH photometric catalog includes extended sources (\texttt{probPSF}=0) brighter than an extinction-corrected Petrosian magnitude of $r=22$ within $180^\prime$ of each cluster center (typically $\gtrsim 7\,R_{200}$). The MACH survey contains SDSS $ugriz$ photometry and adopts foreground-extinction-corrected cModel magnitudes ($r_{\rm cModel,0}$) for subsequent analyses.

The MACH survey derives the stellar mass of each member galaxy by fitting the spectral energy distribution (SED) with the Code Investigating GALaxy Emission (CIGALE) code \citep{Boquien2019}. We fit the extinction-corrected SDSS $ugriz$ cModel photometry with the spectroscopic redshift fixed for each galaxy, and adopt the \citet{Bruzual2003} stellar population synthesis models with a \citet{Chabrier2003} initial mass function. \citet{Park2026b} describe the stellar mass measurement in detail.

The MACH survey provides deep and complete spectroscopy for the target cluster fields. The primary spectroscopy comes from MMT/Hectospec \citep{Fabricant2005}, a 300-fiber spectrograph deployable over a one-degree field of view with wavelength coverage $\sim 3750$--$9100~\angstrom$ at $R \sim 1500$. The MACH catalogs also include publicly available spectroscopy from SDSS DR18, Dark Energy Spectroscopic Instrument (DESI) DR1 \citep{AbdulKarim2025}, and redshifts from the literature (e.g., the NASA/IPAC Extragalactic Database, NED). The MACH survey provides 15572 spectroscopic redshifts within $R_{200}$ and 38638 within $2\,R_{200}$. The spectroscopic completeness for $r_\mathrm{cModel,0}<21.0$ is 87.7\% within $R_{200}$ and 67.9\% within $2\,R_{200}$. The stellar mass measurement of the MACH survey is complete to $10^9\,\Mdot$. 

The dense spectroscopic coverage of the MACH survey enables cluster membership identification and precise dynamical mass estimates. The MACH survey adopts the caustic technique \citep{Kang2024} to define membership. The caustic technique derives the dynamical mass profile as a function of radius from the escape-velocity profile in projected phase space \citep{Diaferio1997, Diaferio1999, Serra2011}. The caustic technique requires $\gtrsim 200$ members per cluster for a reliable mass profile \citep{Serra2011, Pizzardo2023}. \citet{Serra2013} applied the caustic technique to mock redshift surveys, where the true 3D membership of every galaxy is known, and showed that the caustic technique recovers $\sim 95\%$ of the true members within $3\,R_{200}$. The dense MACH spectroscopy comfortably exceeds the sampling requirement for a robust caustic mass profile, yielding 6020 caustic-identified members within $2\,R_{200}$. Among these members, we use 4399 member galaxies with $M_\star > 10^9\,\Mdot$ to measure the stellar mass profile.

\section{Testing the PEEL Model with Simulations} \label{sec:simulation_test}

\subsection{Accuracy Tests of the PEEL Stellar Mass Profile} \label{sec:accuracy_test}

We test the performance of the PEEL model using the TNG-Cluster simulations. In the TNG-Cluster simulation, the true positions and the stellar masses of cluster member subhalos (galaxies) are available. We first construct the true 3D stellar mass profiles of the TNG-Cluster halos based on the 3D distance of member subhalos with respect to the cluster center. 

We also derive the PEEL profile for TNG-Cluster halos within $2\,R_{200}$. We limit the PEEL model within $2\,R_{200}$ to account for the observational limitations of the MACH survey, which provides relatively complete spectroscopic membership of the target clusters (i.e., $\sim 70\%$; \citealp{Park2026b}). We also normalize both PEEL and 3D profiles based on $R_{200}$.

\begin{figure}[tp]
    \centering
    \includegraphics[width=\columnwidth]{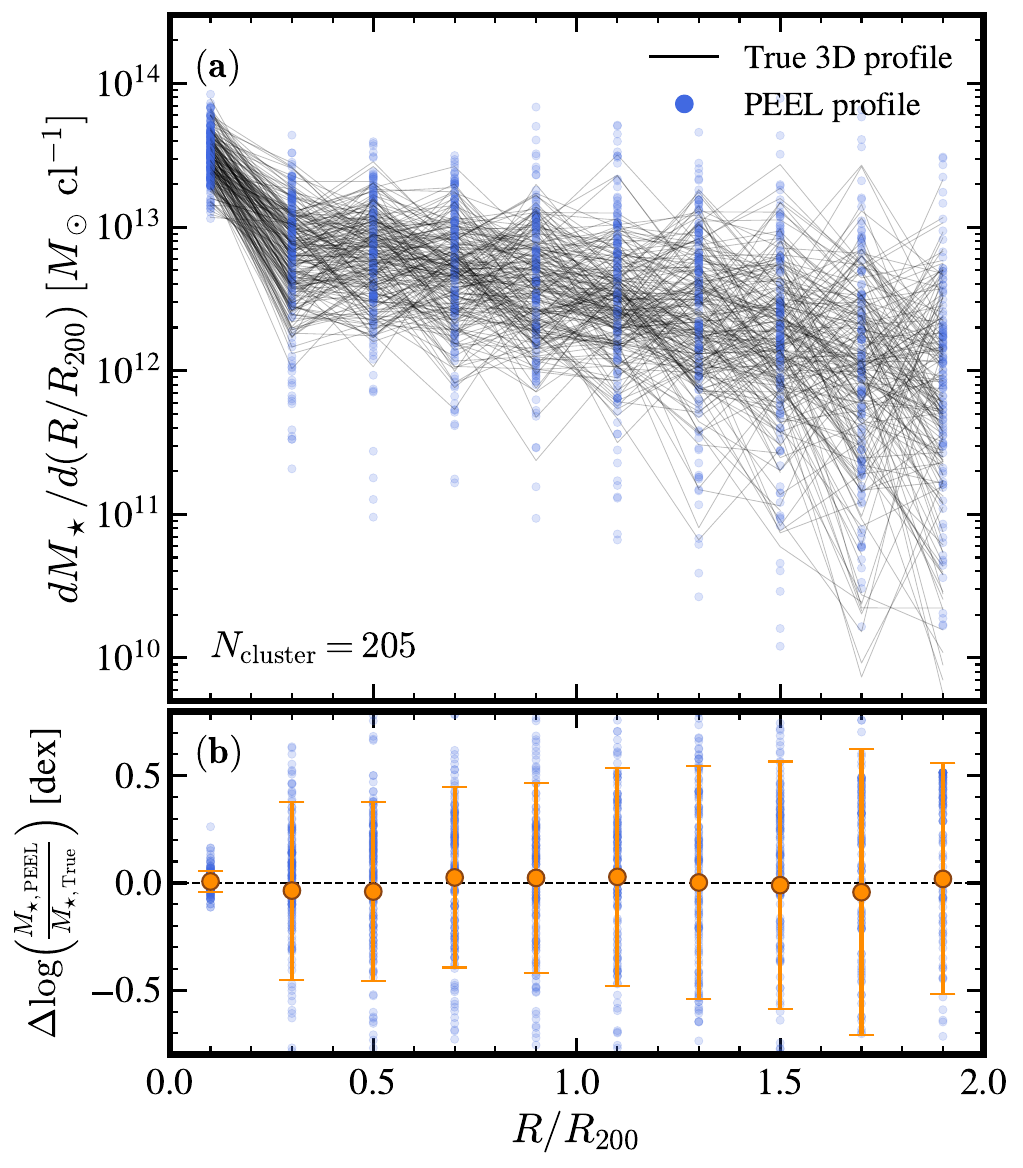}
    \caption{(a) Stellar mass profiles of TNG-Cluster halos within $2\,R_{200}$. Black lines show the true 3D stellar mass profiles. Blue dots show the PEEL profiles based on the deprojection on the $xy$-plane in the simulations. (b) Offsets between the PEEL and true stellar mass profiles as a function of normalized cluster-centric distance. Orange circles indicate the mean and $1\sigma$ scatter. }
    \label{fig:peel_single_halo_test}
\end{figure}

For constructing PEEL profiles of TNG-Cluster halos, we project all cluster halos onto the $xy$-plane of the simulation coordinates. In other words, we simply adopt the $z$-axis as the line-of-sight direction. We then compute $\Sigma^{\star}_{R_{j}}$ to derive the PEEL profile ($\mathcal{M}_\star^{R_i}$) of each halo as $dM_{\star} / d(R/R_{200})$ in units of $\Mdot\,\mathrm{cl}^{-1}$, i.e., stellar mass per cluster per unit $R/R_{200}$. Figure~\ref{fig:peel_single_halo_test}(a) shows the true (black lines) and the PEEL (blue dots) profiles of individual TNG-Cluster halos.

Figure~\ref{fig:peel_single_halo_test}(b) shows the offset between the PEEL and the true 3D stellar mass profiles as a function of normalized cluster-centric radius. Orange circles and error bars indicate the mean offsets and the $1\sigma$ scatters, respectively. The mean offset is consistent with zero, indicating that the PEEL model generally reconstructs the 3D mass profile well. The uncertainty in the offset is $\sim 0.05$ dex at the innermost region, and $0.4$--$0.5$ dex at around $R_{200}$.

\begin{figure*}[tp]
    \centering
    \includegraphics[width=\textwidth]{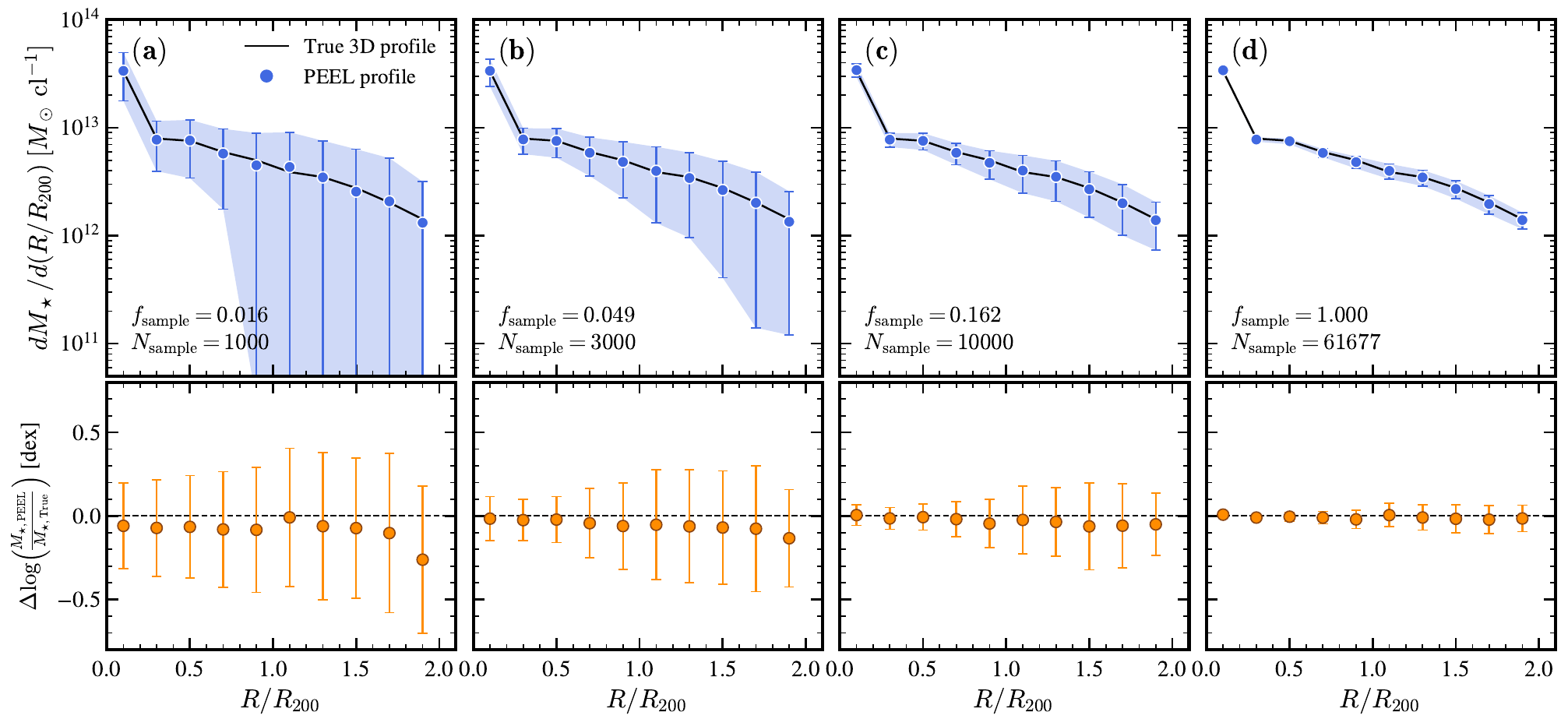}
    \caption{The performance of the PEEL model depending on the size of galaxy samples. From the stacked sample of 205 TNG-Cluster halos, we select 1000, 3000, 10000, and all galaxies to reconstruct the PEEL profile (from left to right). (Upper panels) Black lines show the true mean 3D stellar mass profile per cluster. Blue points and shaded bands show the PEEL mean and $1\sigma$ scatter from 1000 mock observed samples. We normalize the PEEL profiles derived from each subsample based on the number ratio with respect to the total number of the sample. (Bottom panels) The offset between the PEEL and the 3D stellar mass profiles in log scale. }
    \label{fig:peel_galaxy_sampling_test}
\end{figure*}

We also test the performance of the PEEL model depending on the sample size (Figure~\ref{fig:peel_galaxy_sampling_test}). The stacked sample of 205 TNG-Cluster halos consists of 61677 galaxy subhalos. For this test, we randomly select 1000 (1.6\%), 3000 (4.9\%), and 10000 (16.2\%) galaxies and derive the PEEL profile from a random viewing angle. We repeat the random sampling and the PEEL profile derivation 1000 times.
We also derive the PEEL profile from the entire stacked sample by changing the 1000 viewing angles. For the PEEL profiles constructed from subsamples, we normalize the profile with the number ratio with respect to the stacked sample. As in Figure~\ref{fig:peel_single_halo_test}, we compute the offsets between the PEEL profiles of subsamples and the 3D mass profiles in log scale.

Figure~\ref{fig:peel_galaxy_sampling_test} displays the resulting performance of the PEEL profiles. The mean offset and the scatter decrease when we use a subsample with a larger number of galaxies. The scatter is $\sim 0.4$~dex for $N_{\mathrm{sample}} = 1000$, $\sim 0.25$~dex for $N_{\mathrm{sample}} = 3000$, and $\sim 0.15$~dex for $N_{\mathrm{sample}} = 10000$. With the full sample ($N_{\mathrm{sample}} = 61677$), the PEEL model recovers the true 3D profile with a scatter of $\sim 0.05$~dex. 

Our test indicates that stacking multiple clusters with a larger number of member galaxies certainly improves the performance of the PEEL model. A stack of $N$ clusters increases the number of input galaxies for the PEEL model by a factor of $\sim N$. Furthermore, stacking many clusters with various asymmetry and elongation with respect to the projected plane reduces the uncertainties introduced by the departure from the spherical symmetry of individual clusters.

\begin{figure}[tp]
    \centering
    \includegraphics[width=\columnwidth]{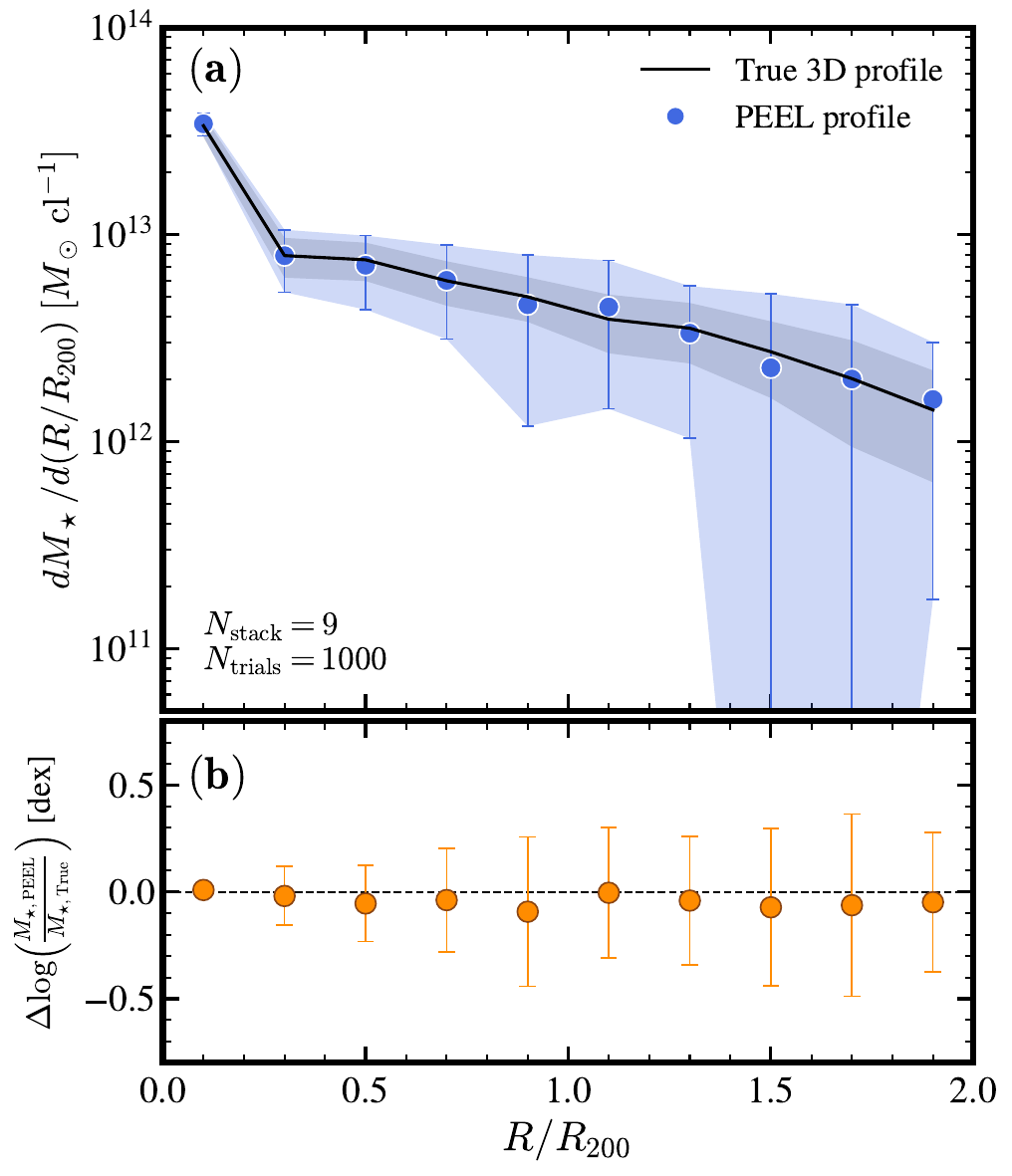}
    \caption{(a) Comparison between the true mass profile and the PEEL profile of the MACH-like simulated mock samples selected from the TNG-Cluster halos. We randomly select nine clusters within the same cluster mass range as the MACH clusters from TNG-Cluster halos and repeat this sampling 1000 times. The black line and blue points show the mean of the true and PEEL profiles derived from the 1000 resamples. The gray and blue shaded regions display the $1\sigma$ scatter of the 1000 true and PEEL profiles, respectively. (b) Offset between the PEEL and the true profiles and the $1\sigma$ scatter of the offset. }
    \label{fig:stack_uncertainty}
\end{figure}

We next examine the expected performance of the PEEL model when applied to the MACH-like survey based on the subsample of TNG-Cluster halos. For this test, we first select clusters within the same mass range as the MACH clusters. We also randomly select nine clusters, the same number of the MACH clusters. We note that the MACH-like mock subsample typically contains $\sim 2700$ subhalos with $M_\star > 10^9\,\Mdot$ within $2\,R_{200}$, much smaller than the number of members (i.e., $\sim 4400$) within the same mass range and cluster-centric radius in the MACH stacked sample. Despite the difference, we still select the same number of clusters to compare the performance depending on the number of clusters. 

Figure~\ref{fig:stack_uncertainty}(a) compares the mean 3D profile (black line) with the mean PEEL profile (blue points) of the MACH-like mock samples. Figure~\ref{fig:stack_uncertainty}(b) shows the mean and $1\sigma$ scatter of the offset between the two profiles. The PEEL profile generally recovers the 3D profile well; averaged over radii, the mean offset is $-0.04$ dex. The scatter is $\sim 0.01$ dex within $R < 0.2\,R_{200}$ and increases to $\sim 0.4$ dex at larger radii. However, the uncertainties quoted here are the conservative ones because the number of galaxies we used here is smaller than the MACH stacked sample. Thus, the actual uncertainties of the PEEL profile of the MACH-stacked sample are expected to be smaller than these quoted values.

\subsection{Source of Uncertainties in the PEEL Model} \label{sec:uncertainties}

The PEEL model recovers the 3D stellar mass profiles for the clusters in the TNG-Cluster simulation, but with some scatter. We examine the sources of uncertainties in reconstructing the 3D mass profile. We first investigate the galaxy-based approach (rather than the particle-based approach) in applying both PEEL and the 3D mass profiles in Section~\ref{sec:bcg_icl_test}. We then examine the impact of uncertainty in the cluster center on the PEEL model results (Section~\ref{sec:centering_offset_test}). Third, we examine the impact of the 3D triaxial structure of the galaxy cluster halos on the PEEL model reconstruction (Section~\ref{sec:halo_triaxiality_test}).

\subsubsection{Subhalo-based versus Particle-based Stellar Mass Profiles} \label{sec:bcg_icl_test}

The PEEL model derives the stellar mass profile by assuming that the stellar mass of each cluster member is concentrated at its center. In other words, we treat the individual cluster members as point sources. This approach is compatible with the analysis of the observational data. Hereafter, we refer to this approach as the subhalo-based approach. The simulation data provides additional information, including the position and the mass of individual stellar particles, which enables tracing the more detailed distribution of stellar particles within cluster halos. We refer to this approach as the particle-based approach.  

We thus examine the performance of the PEEL model based on the particle-based approach to see whether our previous approach based on subhalos reproduces the true mass profile. For testing the particle-based approach, we use all member stellar particles of the FoF group within $2\,R_{200}$ from the cluster center. The particle-based profile differs from the subhalo-based profile in two aspects. First, the particle-based profile places each star particle at its own position rather than at the subhalo center. Second, the particle-based profile includes every stellar particle that is missed by the subhalo-based approach that only uses subhalos with stellar mass larger than $10^{9}~\Mdot$. The stellar particles located within $2\,R_{200}$ but not classified as members of cluster member subhalos with $M_{\star} > 10^{9}~\Mdot$ are included when the particle-based approach is applied.

\begin{figure}[tp]
\centering
\includegraphics[width=\columnwidth]{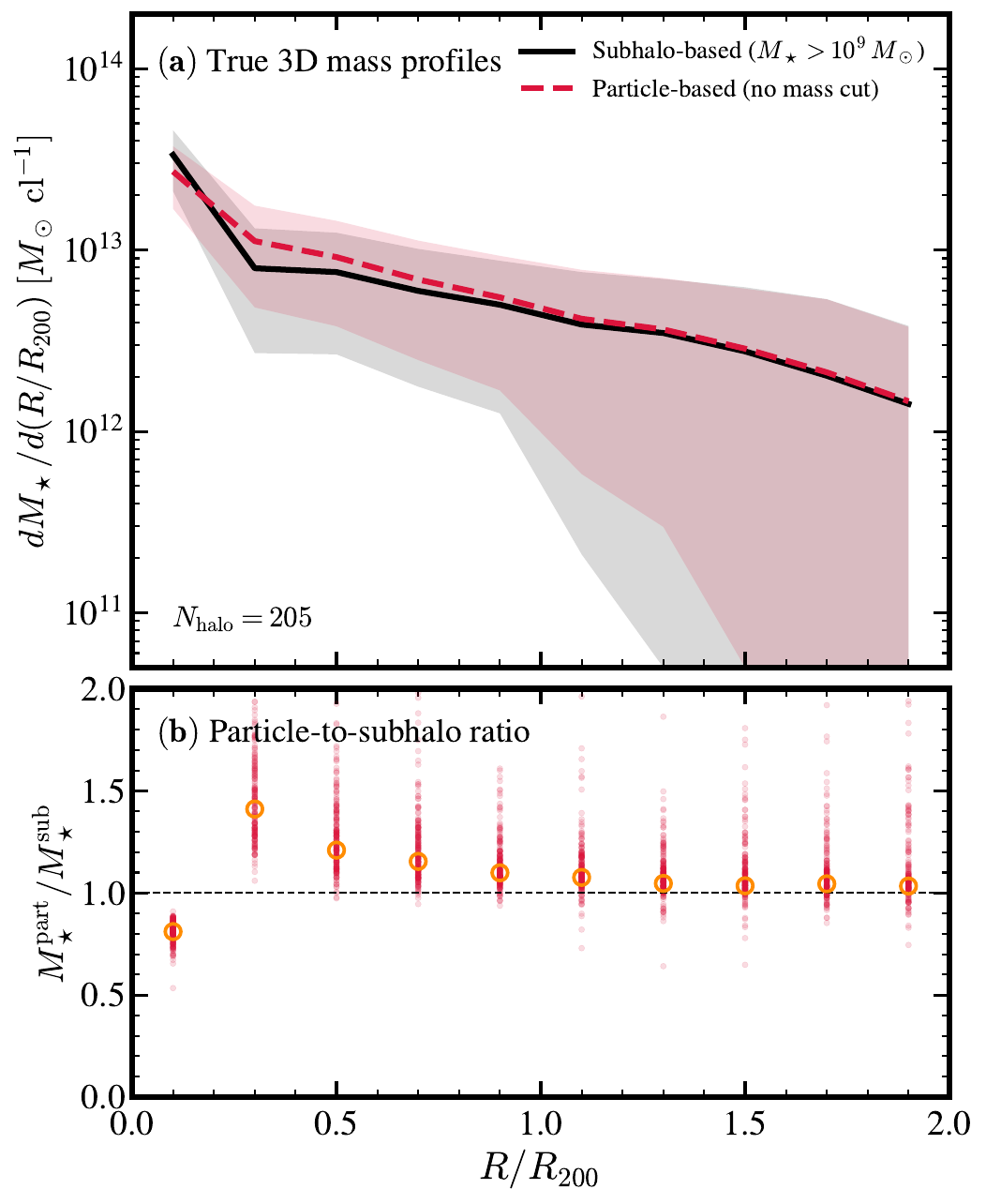}
\caption{Stacked true 3D stellar mass profiles of the TNG-Cluster halos with $10^{14.7}\,\Mdot \le M_{200} \le 10^{15.3}\,\Mdot$, built with the subhalo-based and particle-based estimates. (a)~True 3D stellar mass profiles from the subhalo-based estimate ($M_\star > 10^9\,\Mdot$, black solid) and the particle-based estimate (no mass cut, red dashed); the solid and dashed lines show the mean profiles, and the shaded bands show the $1\sigma$ scatter among the halos. (b)~Ratio of the particle-based to the subhalo-based stellar mass profile $(M_\star^\mathrm{part}/M_\star^\mathrm{sub})$, in each radial bin. Red points show the individual halos, and orange circles show the ratio of the two mean profiles in panel~(a). The dashed line marks a ratio of unity.}
\label{fig:TNGCluster_subhalo_vs_particle_true_profiles}
\end{figure}

Figure~\ref{fig:TNGCluster_subhalo_vs_particle_true_profiles} compares the 3D stellar mass profiles from the subhalo-based (the black solid line) and particle-based (the red dashed line) estimates. The upper panel shows the mean profiles of 205 TNG-Cluster halos, and the lower panel displays the ratio between the two profiles. At the cluster center ($R \lesssim 0.2\,R_{200}$), the particle-based profile is $\sim 19\%$ lower than the subhalo-based profile. The subhalo-based approach counts that the entire stellar mass of the central subhalo are concentrated at the cluster center. On the other hand, the particle-based approach counts stellar particles of central subhalos (as well as other subhalos) at their actual location, regardless of the position of their host subhalos. At intermediate radii ($0.2\,R_{200} \lesssim R \lesssim 0.6\,R_{200}$), the particle-based profile is $\sim 20$--$40\%$ higher than the subhalo-based profile. At $R > 0.6\,R_{200}$, the two profiles agree with each other within $\sim 15\%$. 

The primary difference between the two approaches is the central galaxy and its member particle identification. TNG-Cluster assigns the gravitationally bound stellar particles within the cluster halo that are not members of any satellite subhalo to the central subhalo. In other words, both the member stellar particles of the central galaxy and the intracluster stellar particles are classified as members of the central subhalo. Indeed, the stellar particles of the central subhalo are extended far outside the central region. For the BCG of the TNG-Cluster halos, 78\% of their stellar mass lies within $0.2\,R_{200}$, and 22\% lies beyond $0.2\,R_{200}$. 

\subsubsection{Centering Offset Test} \label{sec:centering_offset_test}

We examine how the performance of the PEEL model depends on the choice of cluster center. Determining the center of a galaxy cluster is not always straightforward. Observational studies adopt various center definitions, including the position of the BCG, the dynamical center, the X-ray peak, and the peak of the weak-lensing significance map, and these centers often do not coincide. We therefore investigate whether the PEEL model yields different results depending on the adopted cluster center.

We quantify the dependence of the PEEL profile on the centering offset. For the TNG-Cluster halos, we derive PEEL profiles using centers displaced from the true center by $\Delta = 0.1$, $0.2$, and $0.4\,R_{200}$ in the projected plane. Here, the true center refers to the position of the central subhalo, i.e., the BCG (Section~\ref{sec:TNG_Cluster}). For each offset center, we derive the PEEL profile 1000 times with randomly chosen viewing angles.

\begin{figure}[tp]
    \centering
    \includegraphics[width=\columnwidth]{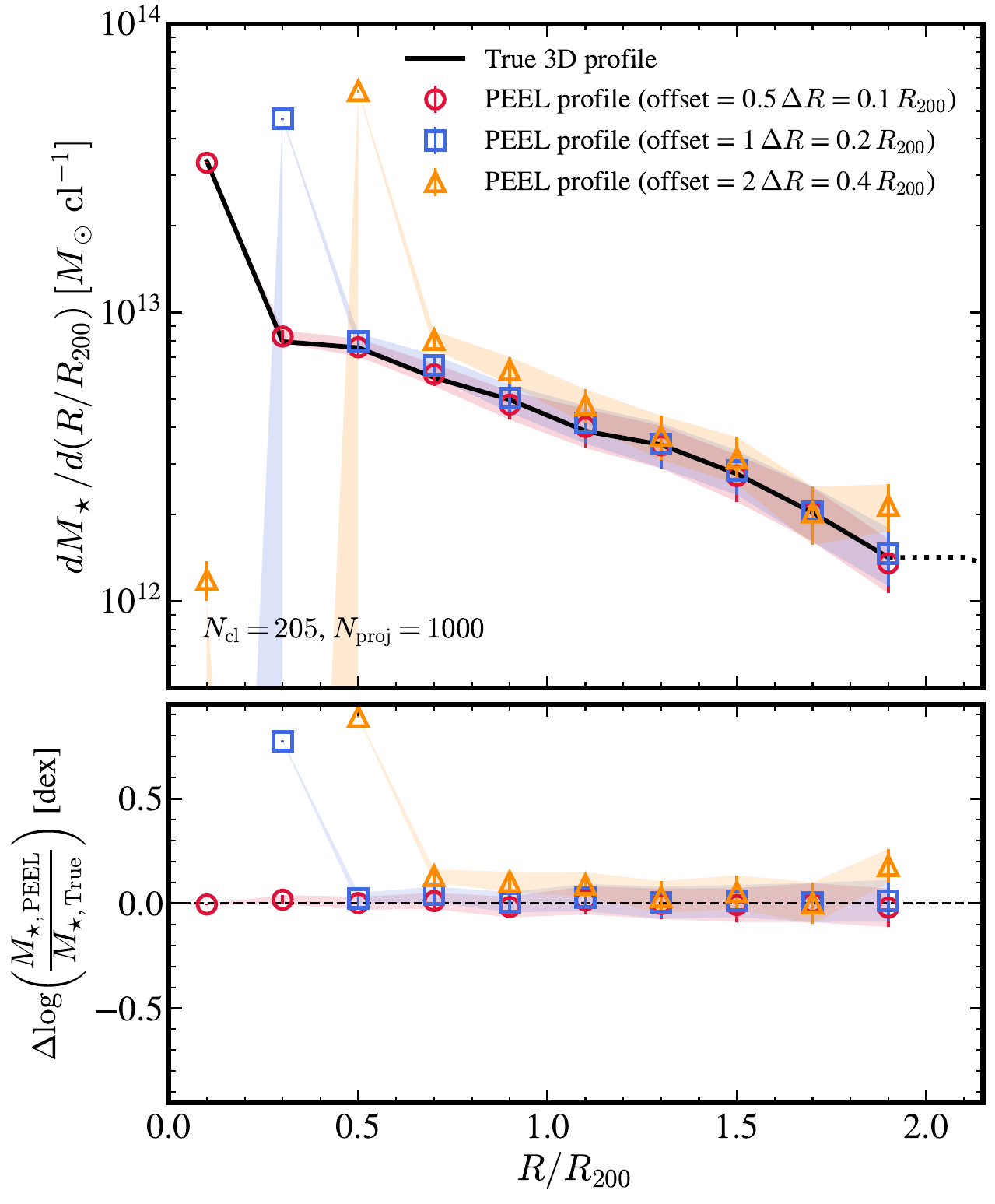}
    \caption{Sensitivity of the PEEL profile to the centering offset, tested on the stacked TNG-Cluster halos with 1000 random projections for each offset. (a) The black solid line shows the true 3D profile; the dotted line shows the true 3D profile beyond $2\,R_{200}$, where we did not apply the PEEL model. Red open circles, blue open squares, and orange open triangles show the mean PEEL profiles for centering offsets of $0.1\,R_{200}$, $0.2\,R_{200}$, and $0.4\,R_{200}$, respectively. The shaded bands indicate the $1\sigma$ scatter among the projections. All PEEL profiles are derived with a radial bin width of $0.2\,R_{200}$. (b) Logarithmic ratio of the PEEL profile to the true 3D profile, $\log~(M_{\star,\mathrm{PEEL}}/M_{\star,\mathrm{True}})$. The horizontal dashed line indicates zero bias. }
    \label{fig:PEEL_center_offset}  
\end{figure}

Figure~\ref{fig:PEEL_center_offset} shows the PEEL profiles derived with the three offset centers. The shaded regions indicate the $1\sigma$ scatter among the 1000 realizations with different viewing angles for a given offset. For the smallest offset ($\Delta = 0.1\,R_{200}$), the PEEL profile agrees with the 3D profile in all radial bins. For larger offsets, the PEEL profiles deviate significantly from the 3D profile: they underestimate the mass density in the core and show spurious peaks at larger radii. In the outer region ($R > R_{200}$), the differences between the PEEL and 3D profiles are insignificant.

These tests show that the PEEL model is sensitive to the centering offset, particularly relative to the radial bin size. The PEEL model recovers the 3D mass profile when the centering offset is smaller than the radial bin size. To test this further, we repeat the analysis with a radial bin size of $0.1\,R_{200}$, which is smaller than the $0.2\,R_{200}$ bin size adopted in Figure~\ref{fig:PEEL_center_offset}. With $0.1\,R_{200}$ bins, a $0.1\,R_{200}$ offset distorts the profile in the same way that a $0.2\,R_{200}$ offset does with $0.2\,R_{200}$ bins, even though the same offset leaves the profile nearly unchanged with $0.2\,R_{200}$ bins. Thus, the sensitivity of the PEEL model to centering errors depends on the size of the offset relative to the radial bin size, and reliable recovery requires a centering offset smaller than the bin size. For the MACH clusters, the mean offset between the dynamical center determined by the caustic technique and the BCG or X-ray peak is typically small ($\sim 0.07\,R_{200}$). Therefore, the PEEL profiles derived with the $0.2\,R_{200}$ bin size adopted here are not strongly affected by centering offsets.

\subsubsection{Halo Triaxiality Test} \label{sec:halo_triaxiality_test}

The PEEL model we build simply assumes a spherical symmetry of a galaxy cluster, but galaxy clusters are often triaxial. In both simulations and observations, there is much evidence for the triaxial shapes of massive clusters (e.g., \citealp{Jing2002, Allgood2006, Bonamigo2015,Limousin2013, Chiu2018a, Sereno2018}). We test how the 3D shape of halos affects the derivation of the PEEL profiles compared to the 3D profile using TNG-Cluster halos. 

We compute the triaxiality of TNG-Cluster halos based on the 3D distribution of member subhalos. We first quantify the mass-weighted inertia tensor of TNG-Cluster halos based on the position and the stellar mass following \citet{Tenneti2014}: 
\begin{equation}
I_{ij} = \frac{\sum_{k=1}^{N} m_k \, x_i^{(k)} x_j^{(k)}}{\sum_{k=1}^{N} m_k}\,,
\label{eq:inertia}
\end{equation}
where $m_{k}$ and $x^{(k)}$ are the stellar mass and the cluster-centric position normalized by $R_{200}$ for $N$ member galaxies. The indices $i,j$ correspond to the $x$-, $y$-, and $z$-axes. The normalization by $\sum_{k} m_{k}$ determines the overall scale. Diagonalizing $I$ yields three eigenvalues: $\lambda_1 \leq \lambda_2 \leq \lambda_3$. The axis ratios of the cluster halo are defined as: 
\begin{equation}
    q_{a} \equiv \frac{c}{a} = \sqrt{\frac{\lambda_1}{\lambda_3}}\,, \qquad
    q_{b} \equiv \frac{b}{a} = \sqrt{\frac{\lambda_2}{\lambda_3}}\,,
    \label{eq:axis_ratios}
\end{equation}
with $0 < q_{a} \leq q_{b} \leq 1$ and the spherical limit $q_{a} = q_{b} = 1$.

We derive PEEL profiles from 1000 random projections for each TNG-Cluster halo; each projection views the same 3D shape along a different line of sight. For a single line-of-sight $j$, we compute the ratio between the PEEL and the 3D profiles as a function of cluster-centric radius $i$: $\delta_i = \log(M_{\star,\mathrm{PEEL}}^{(i)} / M_{\star,\mathrm{True}}^{(i)})$. We then compute the root-mean-square (RMS) of $\delta_{i}$ over the radii: 
\begin{equation}
    \sigma_\mathrm{LoS} = \sqrt{\frac{1}{N_\mathrm{bins}} \sum_{i=1}^{N_\mathrm{bins}} \delta_i^{2}}\,,
    \label{eq:delta_LoS}
\end{equation}
where $N_\mathrm{bins}$ is the number of used bins along the line of sight. For each cluster, we combine $\sigma_\mathrm{LoS}$ of the 1000 random projections in quadrature to obtain a single uncertainty due to the deprojection: 
\begin{equation}
	\sigma_\mathrm{halo} = \sqrt{\frac{1}{N_\mathrm{LoS}} \sum_{j=1}^{N_\mathrm{LoS}} \sigma_{\mathrm{LoS},\,j}^{2}}\,,
    \label{eq:delta_halo}
\end{equation}
where $N_\mathrm{LoS} = 1000$ is the number of random projections. Thus, $\sigma_\mathrm{halo}$ measures the typical deviation (in dex) of the PEEL profile from the true 3D profile of each halo, including both the systematic offset and the scatter.

We exclude radial bins in which the PEEL profile is negative or the 3D profile is empty (i.e., the 3D shell contains no galaxies), because $\delta_{i}$ cannot be computed for these bins. Across the full sample, PEEL returns a negative shell mass in $19.0\%$ of all radial bins, and the true 3D shell is empty in an additional $2.0\%$. The excluded bins lie mostly in the cluster outskirts and contain only $\sim 7\%$ of the total stellar mass and $\sim 15\%$ of the galaxies. Removing them therefore has a negligible impact on $\sigma_\mathrm{halo}$.

\begin{figure}[tp]
    \centering
    \includegraphics[width=\columnwidth]{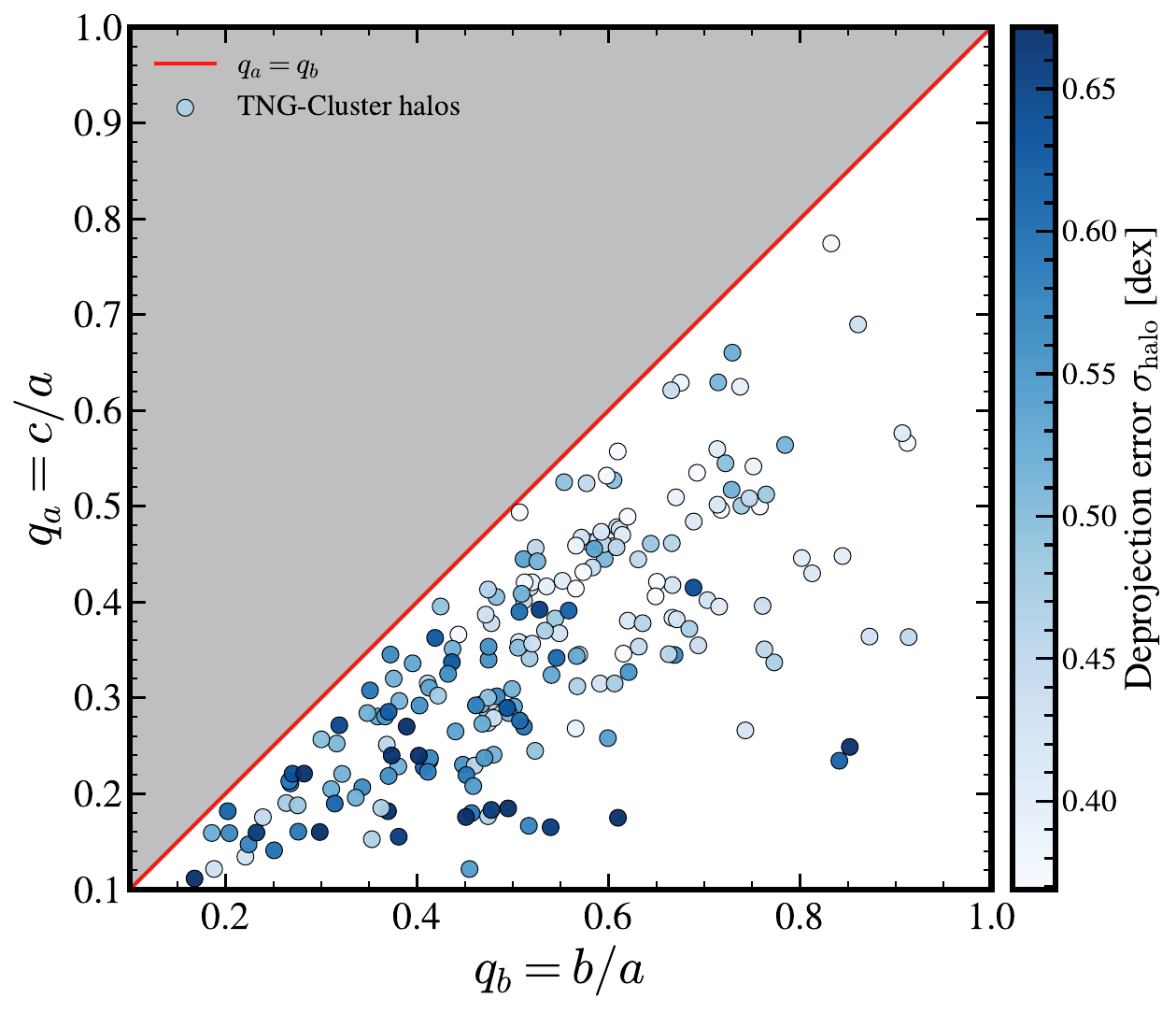}
    \caption{Deprojection error $\sigma_\mathrm{halo}$ for TNG-Cluster halos in the 3D axis-ratio plane $(q_b,\,q_a)$. Each marker is one halo; the color encodes $\sigma_\mathrm{halo}$ (Equation~\ref{eq:delta_halo}), the RMS deprojection error across the 1000 random projections of that halo. The red line marks the diagonal $q_a = q_b$; the gray shaded region above the diagonal ($q_{a} > q_{b}$) is physically forbidden.}
    \label{fig:triaxiality_perhalo}
\end{figure}

Figure~\ref{fig:triaxiality_perhalo} shows the deprojection uncertainty, $\sigma_\mathrm{halo}$, of the TNG-Cluster halos on the $(q_b,\,q_a)$ plane, where darker symbols indicate larger deviations of the PEEL profile from the 3D profile. The $\sigma_\mathrm{halo}$ values range from 0.33 to 0.82~dex, with a median of 0.49~dex (i.e., a factor of $\sim 3$). The deprojection uncertainty increases as $q_{a}$ and $q_{b}$ decrease, indicating that clusters with more aspherical shapes tend to have larger uncertainties. The mean $\sigma_\mathrm{halo}$ decreases from $0.54$~dex for the elongated halos ($q_a < 0.4$) to $0.43$~dex for the rounder halos ($q_{a} \geq 0.4$). The cluster shape therefore affects the recovery of the stellar mass profile with the PEEL model.

\section{Comparison between Observed and Simulated 3D Stellar Mass Profiles} \label{sec:comparison}

We next apply the PEEL model to the MACH cluster sample to reconstruct the 3D stellar mass profile from the observations. We stack the nine MACH clusters to reduce the statistical uncertainties; we stack in $R/R_{200}$-normalized coordinates with $\Delta R = 0.2\,R_{200}$ bins, as for the TNG-Cluster stacked sample.

\subsection{Reconstruction of the 3D Stellar Mass Profile of the Observed Clusters} \label{sec:obs_results}

\begin{figure}[tp]
    \centering
    \includegraphics[width=\columnwidth]{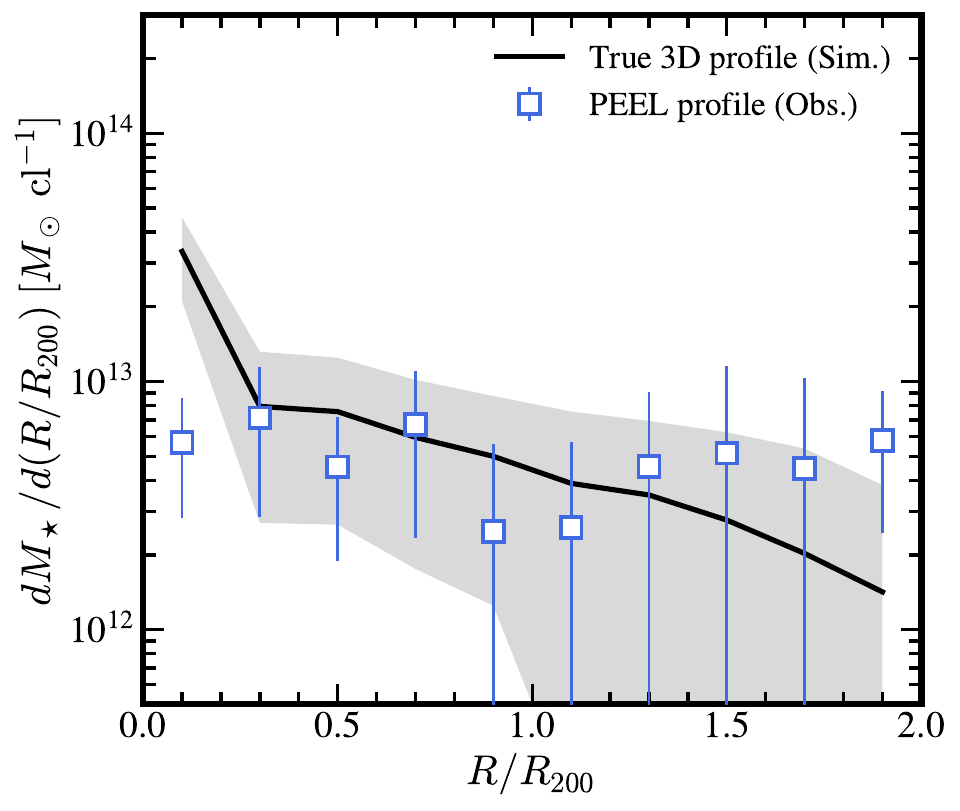}
    \caption{Comparison between the mean 3D stellar mass profile of the TNG-Cluster halos (black solid line) and the PEEL profile derived from the stacked sample of nine MACH clusters (blue squares) for clusters with $10^{14.7}\,\Mdot \leq M_{200} \leq 10^{15.3}\,\Mdot$. The gray shaded band shows the $1\sigma$ scatter of the true 3D profiles of the TNG-Cluster halos. The blue error bars show the $1\sigma$ scatter among the PEEL profiles of the nine individual MACH clusters.}
    \label{fig:PEEL_obssim_comparison}
\end{figure}

We first derive the 2D projected stellar mass profile of the MACH clusters to apply the PEEL model. From the MACH stacked sample, we sum the stellar mass of spectroscopically identified member galaxies at various projected annuli normalized by $R_{200}$ ($\Sigma^\star_{R_i}$); hereafter, $R_\mathrm{proj}$ denotes the cluster-centric radius in the projected observed plane. Because MACH is not absolutely complete, there must be tentative member galaxies that are not included in our galaxy sample, but contribute to the stellar mass profile. We correct the contribution of the missing spectroscopic members following the technique used for correcting stellar mass function (SMF) measurements of the MACH clusters. 

\citet{Park2026b} derived the SMF of the cluster galaxies based on the MACH spectroscopy. For counting the missing cluster members due to lack of spectroscopy, they empirically corrected the SMF. They first empirically derived the member fraction among spectroscopic sample depending on the absolute magnitude, color, and cluster-centric distance. Then, they counted the tentative number of members among the galaxies without spectroscopy in various absolute magnitude, color, and cluster-centric distance bins. The stellar mass of these tentative member galaxies was also estimated based on CIGALE by assuming that these galaxies are at the cluster redshift. 

We follow this approach to identify tentative member galaxies with $M_{\star} > 10^{9}~\Mdot$, and derive the PEEL profile by including the tentative member galaxies. We repeat the tentative member selection 1000 times to take the mean contribution of the statistically selected samples. At $R_\mathrm{proj} < R_{200}$ ($2\,R_{200}$), the number fraction of tentative members is $\sim 13\%$ ($\sim 25\%$) of the total number of members. Because the spectroscopic survey is relatively complete at $R_\mathrm{proj} \lesssim 1.4\,R_{200}$, the membership correction stays at the few-percent level per radial bin there. In the outer region ($R_\mathrm{proj} > 1.4\,R_{200}$), the tentative members outnumber the spectroscopic members, and the correction rises to $\sim 14$--$21\%$ per radial bin; the reconstructed stellar mass profile thus carries a larger uncertainty in the cluster outskirts.

The mean stellar contribution of the tentative cluster members is only $\sim 1.7\%$ of the total stellar mass at $R_\mathrm{proj} < R_{200}$; the contribution increases slightly to $\sim 4.4\%$ at $R_\mathrm{proj} < 2\,R_{200}$ due to the relatively lower spectroscopic completeness in the cluster outskirts. In other words, the member galaxies excluded from our spectroscopic sample (mainly due to their low luminosity) contribute relatively little to the total stellar mass and the stellar mass profile. The stellar mass contribution of the tentative members is much smaller than expected based on their numbers because the tentative members are generally less massive compared to the bright spectroscopically identified members.

We additionally estimate the contribution for the tentative members at $M_{\star} < 10^{9}~\Mdot$ based on the extrapolation of the SMF. Although MACH is one of the densest spectroscopic surveys for local clusters, the completeness is very low at $M_{\star} < 10^{9}~\Mdot$ \citep{Park2026b}. Thus, identification of tentative members based on the incomplete spectroscopic sample may introduce large uncertainty. Instead, we simply compute the total stellar mass at $M_{\star} < 10^{9}~\Mdot$ by integrating the best-fit Schechter function for the MACH SMFs at the same mass range (Appendix~\ref{app:schechter_integration}). \citet{Park2026b} derived the best-fit Schechter function for the MACH SMFs measured at three projected radius bins. We thus compute the total stellar mass of galaxies with $M_{\star} < 10^{9}~\Mdot$ at three radius bins. 

The stellar mass contributions by the low mass galaxies with $M_{\star} < 10^{9}~\Mdot$ are only $2.0\%$ at $R_\mathrm{proj} < 0.5\,R_{200}$, $4.0\%$ at $0.5\,R_{200} \leq R_\mathrm{proj} < 1.0\,R_{200}$, and $3.7\%$ at $1.0\,R_{200} \leq R_\mathrm{proj} < 2.0\,R_{200}$, respectively. Thus, the impact of these low mass galaxies in the determination of the PEEL profile seems relatively insignificant. We do not include these low mass galaxies in the stellar mass profile.

\subsection{Comparison of the Observed and Simulated Stellar Mass Profiles} \label{sec:sm_comparison}

\begin{figure*}[tp]
    \centering
    \includegraphics[width=\textwidth]{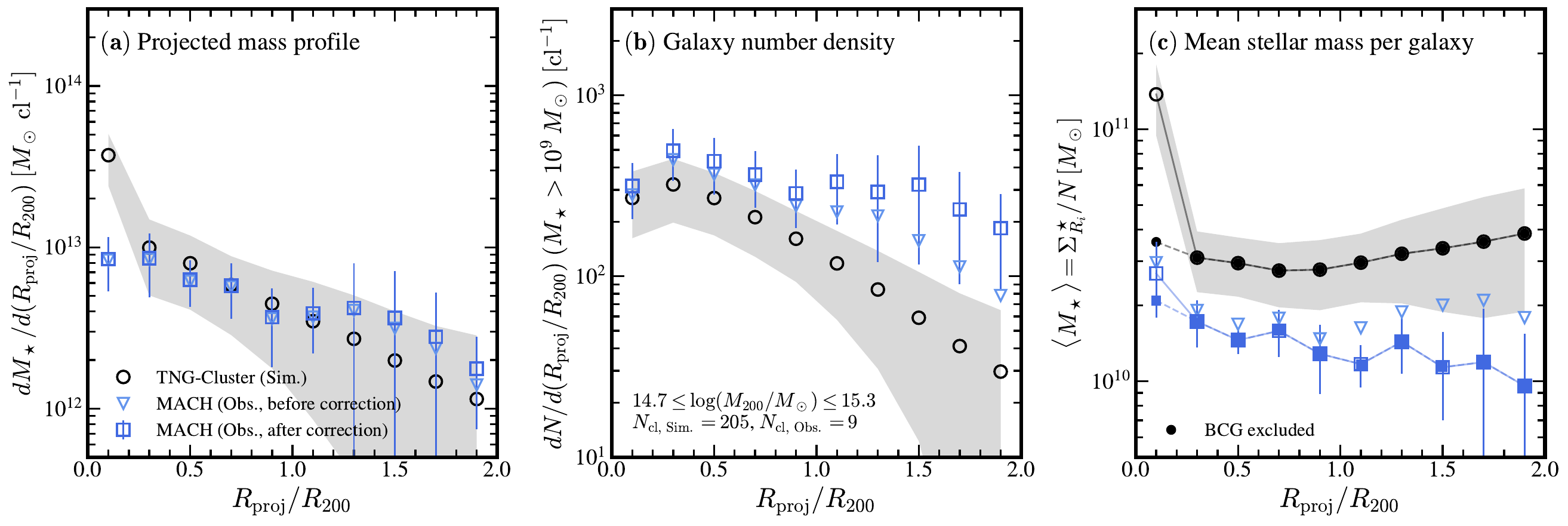}
    \caption{(a)~Projected stellar mass profile, (b)~galaxy number count profile, and (c)~mean stellar mass per galaxy profile of the TNG-Cluster halos and the nine MACH clusters with $10^{14.7}\,\Mdot \leq M_{200} \leq 10^{15.3}\,\Mdot$. Black open circles show the mean of the simulated halos; gray shaded bands indicate the $1\sigma$ scatter among the TNG-Cluster halos. Light-blue open triangles and blue open squares show the MACH profiles before and after the membership correction, respectively; blue error bars indicate the $1\sigma$ scatter among the nine clusters. In panel~(c), the symbols plot the ratio of the mean profiles in panels~(a) and (b); the band and error bars indicate the scatter of the per-cluster $\langle M_\star \rangle$. Filled symbols show the same profiles with the BCGs excluded.}
    \label{fig:obssim_galcount_comparison}
\end{figure*}

\begin{figure*}[tp]
    \centering
    \includegraphics[width=\textwidth]{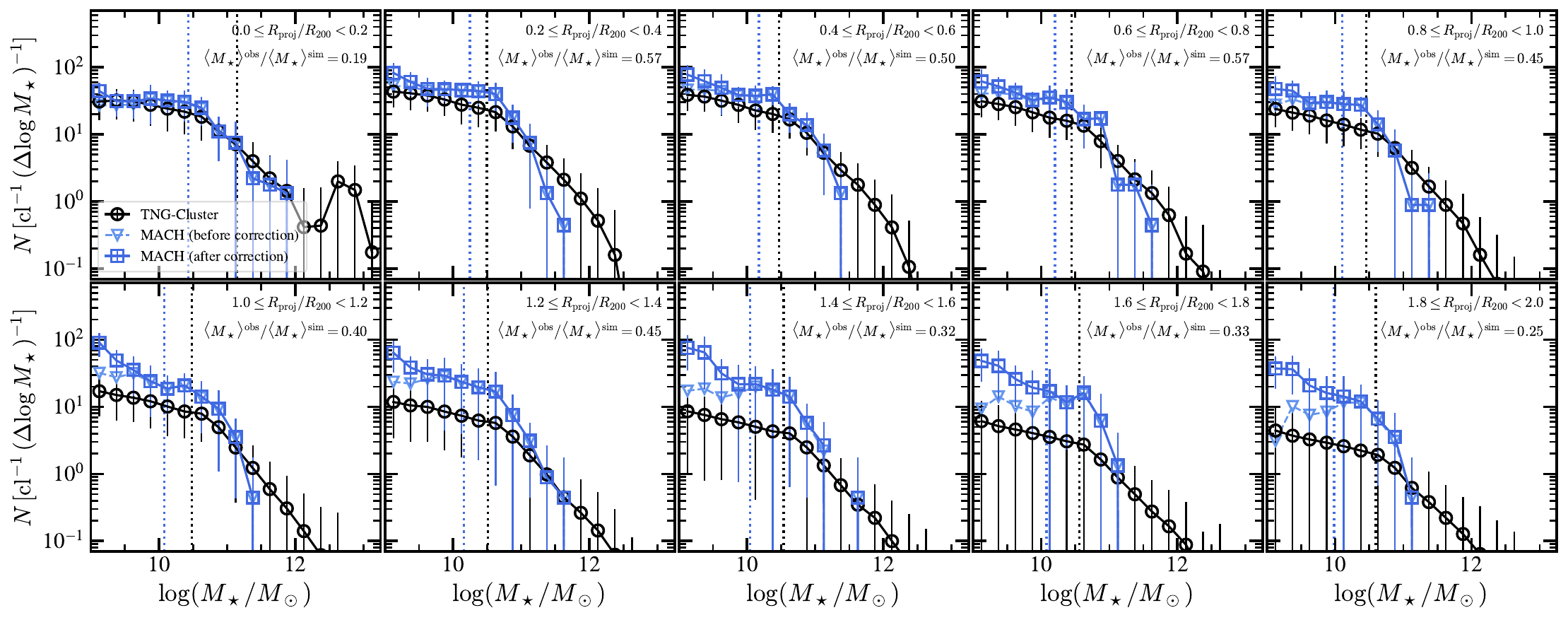}
    \caption{SMFs of the cluster galaxies in 10 projected radial bins of width $0.2\,R_{200}$ for the TNG-Cluster halos and the nine MACH clusters with $10^{14.7}\,\Mdot \leq M_{200} \leq 10^{15.3}\,\Mdot$. Black open circles show the mean of the simulated halos; light-blue open triangles and blue open squares show the MACH SMFs before and after the membership correction, respectively. Error bars indicate the $1\sigma$ scatter among the TNG-Cluster halos and among the nine MACH clusters, as in Figure~\ref{fig:obssim_galcount_comparison}. Black and blue vertical dotted lines mark the mean stellar mass per galaxy $\langle M_{\star} \rangle$ of the simulated and observed samples, respectively, the same values as in panel (c) of Figure~\ref{fig:obssim_galcount_comparison}.}
    \label{fig:obssim_radial_smf}
\end{figure*}

Figure~\ref{fig:PEEL_obssim_comparison} shows the PEEL profile (blue squares) derived from the stacked MACH sample. The mean stellar mass profile fluctuates within $(2.5$--$7.1) \times 10^{12}~\Mdot~\mathrm{cl}^{-1}$ and shows only a weak radial dependence at $R < 2\,R_{200}$. To estimate the uncertainties of the PEEL profile, we also derive the PEEL profile of each MACH cluster after applying the missing member correction. The error bars indicate the $1\sigma$ scatter among these nine single-cluster PEEL profiles.

We compare the PEEL profile of the stacked MACH sample with the mean 3D stellar mass profile of the stacked TNG-Cluster sample (hereafter the simulated profile). The two profiles differ in shape. The simulated profile declines continuously with increasing radius, whereas the observed profile changes little at large radii. Because of this difference in slope, the observed profile exceeds the simulated profile by a factor of $\sim 2$--$4$ at $R > 1.5\,R_{200}$. The largest difference appears in the cluster core, where the simulated profile exceeds the observed profile by a factor of $\sim 5$--$6$ within $0.2\,R_{200}$.

Because the PEEL model reconstructs the 3D profile by removing the contributions of the outer shells from the inner annuli, the difference between the observed and simulated profiles should be traceable to the galaxy distributions in the projected plane. Figure~\ref{fig:obssim_galcount_comparison}(a) compares the observed and simulated projected stellar mass profiles. For the simulated clusters, we derive the projected profile of each TNG-Cluster halo from 1000 random viewing angles and then compute the mean projected profile over all halos (black circles). The gray shaded region indicates the $1\sigma$ scatter among the projected profiles of the TNG-Cluster halos. For the observed clusters, we show the projected stellar mass profiles before (blue triangles) and after (blue squares) the missing member correction, and the error bars indicate the $1\sigma$ scatter among the nine MACH clusters. As in the comparison between the PEEL and 3D profiles, the observed projected profile is flatter than the simulated projected profile.

Figure~\ref{fig:obssim_galcount_comparison}(b) compares the projected galaxy number counts of the MACH and TNG-Cluster clusters, using the same symbols as in panel (a). The number counts reveal two notable features. First, the observed and simulated clusters contain similar numbers of galaxies in the core, indicating that the difference in the projected stellar mass profile in the core does not arise from a difference in galaxy number. Second, the observed clusters generally contain more galaxies than the simulated clusters. In particular, the number of members at $R_\mathrm{proj} > R_{200}$ is $\sim 2$--$3$ times larger in the observed clusters than their counterparts in the simulation. The number difference becomes even larger when we consider the missing member correction (by a factor of $3$--$6$). A larger number of galaxies in the outskirts lowers the central PEEL profile, because the PEEL model removes the contribution of the outer shells first. However, this effect alone cannot account for the factor of $\sim 5$--$6$ difference in the core, given the similar number counts there. Figure~\ref{fig:obssim_galcount_comparison}(c) instead shows that the simulated cluster galaxies have a much higher mean stellar mass per galaxy than the observed cluster galaxies.

Finally, to understand the difference in the mean stellar mass per galaxy, we examine the SMFs of the observed and simulated clusters as a function of cluster-centric radius. Figure~\ref{fig:obssim_radial_smf} shows the SMFs in ten projected annuli. The simulated clusters contain more massive galaxies ($M_{\star} > 10^{11}~\Mdot$) than the observed clusters at all radii. In particular, extremely massive galaxies with $\log (M_{\star}/\Mdot) > 12$ appear only in the cores of the simulated clusters. At $R_\mathrm{proj} < R_{200}$, the total stellar mass of galaxies with $M_{\star} > 10^{11}~\Mdot$ is $\sim 6$ times larger in the simulated clusters than in the observed clusters. This large excess of massive galaxies in the simulated cores drives the significant difference between the observed and simulated stellar mass profiles in the central region.

In contrast, the observed clusters contain more galaxies with $M_{\star} < 10^{11}~\Mdot$ than the simulated clusters, and this difference increases toward the cluster outskirts. At $R_\mathrm{proj} < R_{200}$, the total stellar mass of galaxies with $10^{9}~\Mdot < M_{\star} < 10^{10.5}~\Mdot$ is $\sim 1.7$ times larger in the observed clusters than in the simulated clusters. These intermediate-mass galaxies contribute to the flatter stellar mass profile of the observed clusters.

Our analysis of the radially resolved SMFs indicates that the differences between the observed and simulated stellar mass profiles originate from intrinsic differences in the galaxy populations rather than from the PEEL methodology. Because MACH provides nearly complete spectroscopy for the observed clusters, survey incompleteness cannot explain the flatter observed profile. Incompleteness would reduce the number of observed members, whereas the observed clusters contain more intermediate-mass galaxies than the simulated clusters. The differences between the observed and simulated stellar mass profiles therefore provide an important test for future simulations of galaxy clusters.

\subsection{Stellar Mass Fraction}

\begin{figure*}[tp]
    \centering
    \includegraphics[width=\textwidth]{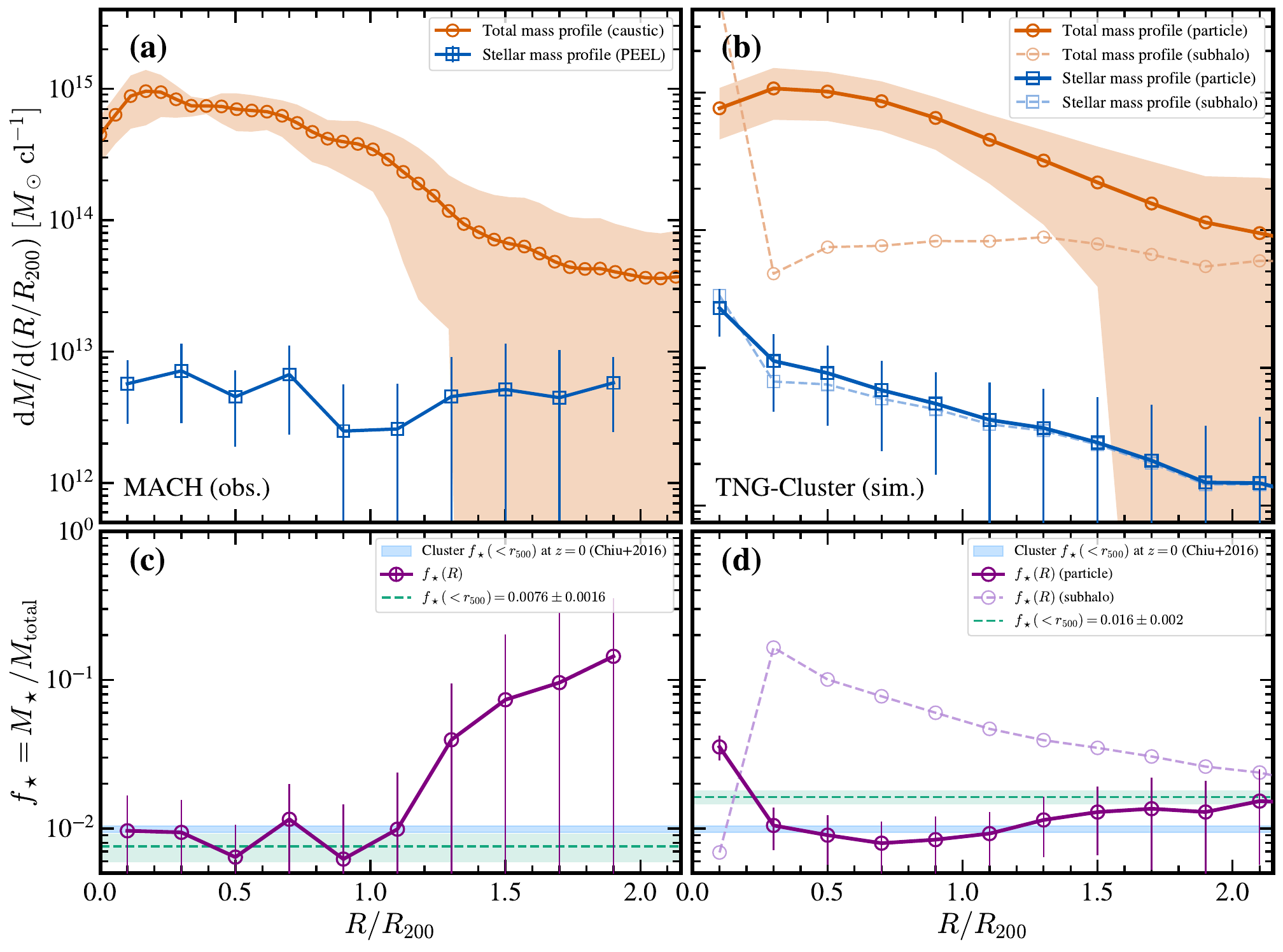}
    \caption{Radial mass profiles and stellar mass fractions for (left panels) the stacked MACH cluster sample (nine clusters, $10^{14.7}~\Mdot \leq M_{200} \leq 10^{15.3}~\Mdot$) and (right panels) the TNG-Cluster simulation (205 halos, same mass range).
    (a) Observed radial mass profiles $\mathrm{d}M/\mathrm{d}(R/R_{200})$ in units of $\Mdot\,\mathrm{cl}^{-1}$. Orange circles show the total dynamical mass from the caustic technique, and blue squares show the PEEL profile; the shaded region and the error bars indicate the $1\sigma$ scatter among total mass and the stellar mass profiles of the MACH clusters.
    (b) Simulated radial mass profiles. Orange circles show the total mass and blue squares show the stellar mass. In each pair, the solid line in the darker color marks the particle-based profile, which we adopt as the main measurement, and the dashed line in the lighter color marks the subhalo-based profile. The particle-based stellar mass includes the diffuse stellar component. The shaded region and the error bars indicate the $1\sigma$ scatter among the particle-based profiles of TNG-Cluster halos. 
    (c) Observed stellar mass fraction $f_\star = M_\star/M_\mathrm{total}$ as a function of $R/R_{200}$ (purple circles).
    (d) Simulated stellar mass fraction. Dark purple solid and light purple dashed lines show the particle-based and subhalo-based $f_\star$, respectively. 
    In panels (c) and (d), the blue shaded band marks $f_{\star} = 0.0099 \pm 0.0005$, derived from the $f_{\star}$--$M_{500}$ scaling relation of \citet{Chiu2016a} for a cluster with $M_{500} = 6 \times 10^{14}~\Mdot$ at $z = 0$. The green dashed lines and shaded regions show the mean and $1\sigma$ scatter of $f_{\star}$ within $R_{500}$ for the MACH clusters and the TNG-Cluster halos, respectively. }
    \label{fig:baryon_fraction}
\end{figure*}

The stellar mass fraction with respect to the total mass of a galaxy cluster (i.e., $f_{\star}$) measures the integrated efficiency of baryons into stars in the target clusters (e.g., \citealp{Lin2003, Kravtsov2012}). Observationally, many studies report the $f_{\star}$ measured within $R_{500}$ from the summed stellar mass of the member galaxies. \citet{Lin2003} measure the sum of the member stellar masses based on the $K$-band luminosity function. \citet{Gonzalez2013} apply a similar calculation but improve the calculation of the stellar mass fraction by including the intracluster light (ICL) contribution. \citet{Chiu2016a, Chiu2018b} derive the stellar masses of individual member galaxies from SED fitting, but without the ICL contribution. \citet{Andreon2010} computes the total stellar mass based on the SDSS luminosity function integration. The $f_{\star}$ varies depending on the cluster halo mass: $\sim 3.0\%$ for clusters with $M_{500} \sim 5 \times 10^{13}~\Mdot$ and $\sim 0.5\%$ for clusters with $M_{500} \sim 2 \times 10^{15}~\Mdot$ \citep{Chiu2018b} with no significant redshift dependence \citep{Chiu2016b, Chiu2018b}. 

Here, we derive the $f_{\star}$ of observed and simulated clusters based on the PEEL profile and the true 3D stellar mass profile, respectively. In addition to the single value of $f_{\star}$ at the fixed aperture, we also examine the radial dependence of $f_{\star}$. 

For the MACH clusters, we use the total mass profile derived by the caustic technique \citep{Diaferio1997, Diaferio1999, Serra2011}. The caustic technique computes the dynamical mass profile based on the line-of-sight velocity distribution of cluster members (Section~\ref{sec:MACH_survey}). We simply divide the PEEL profile by the caustic mass profile to obtain the $f_{\star}$ profile. We also note that we compute the PEEL and the total mass profiles based on the stacked sample of nine MACH clusters. The shaded region for the total mass profile and the error bars for the stellar mass profile indicate the $1\sigma$ scatter among nine MACH clusters. We propagate the $1\sigma$ scatter in the total and the stellar mass profiles to compute the error bars for $f_{\star}$. For the TNG-Cluster halos, the shaded region (total mass) and the error bars (stellar mass and $f_\star$) show the $1\sigma$ scatter of the particle-based profiles among the TNG-Cluster halos.

We also compute the $f_{\star}$ profile for TNG-Cluster halos based on the true 3D stellar mass profile divided by the total mass distribution including the dark matter and non-stellar components (i.e., gas, black holes). We also use both the subhalo-based and particle-based approaches (Section~\ref{sec:bcg_icl_test}). In the case of the subhalo-based technique, we obtain both stellar mass and total mass of individual member subhalos and compute the $f_{\star}$ profile based on the subhalos at different cluster-centric radii. For the particle-based approach, we count all stellar and other particles of the FoF group of each halo at different cluster-centric radii and compute the $f_{\star}$ profile. For the TNG-Cluster halos, we use the true 3D stellar mass profile, not the PEEL profile, because the 3D positions of the particles and subhalos are known.

Figure~\ref{fig:baryon_fraction} displays the total and the stellar mass profiles for observed and simulated clusters (the upper panels). The total mass profiles of observed and simulated clusters have similar shapes within $R_{200}$. In the case of the subhalo-based approach, the particles (specifically, the dark matter particles) extended over the wider region are classified based on the member of the central galaxies, and thus the total mass profile shows a stronger concentration at the core and deficit in the outer region. The PEEL stellar mass profile of the observed clusters is much flatter than that of the simulated clusters (see Section~\ref{sec:sm_comparison} for more details). 

The lower panels of Figure~\ref{fig:baryon_fraction} show the $f_{\star}(R)$ of the observed and simulated clusters. The $f_{\star} (R)$ we derived from MACH clusters remains almost constant at $0.009 \pm 0.002$ within $R_{200}$. The $f_{\star}$ of the MACH clusters within $R_{500}$ is $0.0076 \pm 0.0016$, where the uncertainty indicates the $1\sigma$ scatter of the nine clusters. At $R > R_{200}$, $f_{\star} (R)$ increases drastically up to $\sim 0.1$, but the uncertainty of the $f_{\star}$ is also extreme due to the large uncertainty in the caustic mass. For comparison, we mark the $f_{\star}$ at $z = 0$ (i.e., $0.0099 \pm 0.0005$) derived from the $f_{\star}$--$M_{500}$ scaling relation for a cluster with a mass $M_{500} =  6 \times 10^{14}\,\Mdot$ \citep{Chiu2016a}, which is consistent with the $f_{\star}$ of MACH clusters at $R < R_{200}$. 

The right panels of Figure~\ref{fig:baryon_fraction} show similar mass profiles for the TNG-Cluster halos. We plot the mass profiles based on the subhalo-based (dashed lines) and the particle-based (solid lines) approaches together. We use the particle-based profiles for the interpretation because they sum all stellar particles without relying on the subhalo assignment and thus provide a complete census of the stellar mass. The stellar mass profiles of two approaches do not differ significantly. 

At $R < R_{200}$, the $f_{\star} (R)$ of the TNG-Cluster halos is generally consistent with the values we derived from the MACH clusters and the values expected from the scaling relation. Particularly, the $f_{\star} (R)$ at $0.2\,R_{200} < R < 1.0\,R_{200}$ for the TNG-Cluster halos is $0.009 \pm 0.001$. In TNG-Cluster, the $f_{\star}$ varies little at the larger radii because both total mass and stellar mass profiles decrease continuously as the radius increases. 

We also note that the $f_{\star}$ for the TNG-Cluster halos at the cluster core is much larger than the observed value (by a factor of four). This excess mainly results from the immensely large stellar mass of the central galaxies in the simulated clusters (see Section~\ref{sec:sm_comparison}). Because of the larger $f_{\star}$ at the core, the mean $f_{\star}$ we derived within $R_{500}$ (the green horizontal line) is also larger than the observed value. The comparisons between the $f_{\star}$ in observations and simulations suggest further tests with well-sampled clusters that enable the derivation of the total and the stellar mass profile measurements. 

The PEEL profile provides a platform to translate the projected mass profile to the 3D mass profile, which also enables the estimation of the total mass density within the clusters. Further tests with a larger sample of clusters with dense spectroscopy certainly reduce the uncertainties in the reconstructed stellar mass profiles and resulting stellar mass fraction profiles.

\section{Conclusion} \label{sec:conclusion}

We apply the PEEL (Projection-to-Enclosed Estimator by Layers) model to recover
3D stellar mass profiles from 2D projected mass profiles under the assumptions of spherical symmetry and a void field beyond the outermost shell. The MACH survey, a dense spectroscopic sample of nine massive clusters at $0.07 < z < 0.11$, provides the projected stellar mass profile; the caustic technique provides the total mass profile from the same spectroscopic redshifts. Based on the two profiles, we measure the stellar mass fraction as a 3D radial profile, $f_\star(R)$, within $2\,R_{200}$. The spectroscopic membership keeps the correction for missing members small; the tentative members add only $\sim 1.7\%$ of the stellar mass at $R_\mathrm{proj} < R_{200}$ and $\sim 4.4\%$ at $R_\mathrm{proj} < 2\,R_{200}$.

We test the PEEL model on 205 TNG-Cluster halos at $z \approx 0.07$ with $10^{14.7}\,\Mdot \leq M_{200} \leq 10^{15.3}\,\Mdot$. The PEEL model recovers the true 3D stellar mass profile without bias when applied to stacked samples. For a nine-cluster stack matched to the MACH sample size, the mean offset averaged over radii is $-0.04$~dex; the scatter is $\sim 0.01$~dex in the core and $\sim 0.4$~dex in the outskirts.

We also examine three sources of uncertainty in reconstructing the 3D mass profile. Switching from the subhalo-based to the particle-based stellar mass assignment changes the simulated profile by at most $\sim 40\%$, with a $\sim 19\%$ reduction in the core. Centering offsets smaller than the radial bin width of $0.2\,R_{200}$, well above the MACH caustic--BCG/X-ray offset of $\sim 0.07\,R_{200}$, leave the profile nearly unchanged. The deprojection error is smaller for rounder halos.

We compare the 3D stellar mass profile of the MACH clusters with the TNG-Cluster stack in the same mass range. The simulated profile exceeds the observed profile by a factor of $\sim 5$--$6$ within $0.2\,R_{200}$ and falls below the observed profile by a factor of $\sim 2$--$4$ beyond $1.5\,R_{200}$. The central excess reflects the high stellar mass of the simulated central galaxy rather than a difference in galaxy count. In contrast, the deficit in the outskirts reflects fewer galaxies in the simulation; at $R_\mathrm{proj} > R_{200}$, the observed clusters contain $\sim 3$--$6$ times more galaxies than the simulated clusters after the missing-member correction. Above $10^{11}\,\Mdot$, the total stellar mass in galaxies is $\sim 6$ times larger in the simulated clusters at $R_\mathrm{proj} < R_{200}$; below $10^{11}\,\Mdot$, the observed clusters contain more galaxies, and the difference grows toward the outskirts.

We derive $f_\star(R)$ of the MACH clusters from the PEEL and caustic profiles and compare it with the particle-based $f_\star(R)$ from TNG-Cluster. The $f_\star(R)$ of the MACH clusters remains almost constant at $0.009 \pm 0.002$ within $R_{200}$, consistent with the $f_\star$--$M_{500}$ scaling relation of \citet{Chiu2016a}; the $f_\star$ within $R_{500}$ is $0.0076 \pm 0.0016$. At $R > R_{200}$, $f_\star(R)$ increases up to $\sim 0.1$, but its uncertainty is also large because of the large uncertainty in the caustic mass.

Within $R_{200}$, the $f_\star(R)$ of the TNG-Cluster halos is generally consistent with that of the MACH clusters; at $0.2\,R_{200} < R < 1.0\,R_{200}$, the simulated $f_\star(R)$ is $0.009 \pm 0.001$. In the cluster core, the simulated $f_\star$ is about four times the observed value because of the large stellar mass of the central galaxies, and the simulated $f_\star$ within $R_{500}$ is thus also larger than the observed value. Compared with the MACH clusters, TNG-Cluster concentrates more stellar mass in the cluster core and contains less stellar mass in the outskirts; the radial dependence of $f_\star$ reveals this difference, which the single value of $f_{\star}$ at the fixed aperture does not.

\section{Acknowledgements}
We thank Ken Rines for his insightful suggestions that improved this manuscript. This work was supported by the National Research Foundation of Korea (NRF) grant funded by the Korea government (MSIT; RS-2026-25488355). This work was supported by Creative-Pioneering Researchers Program through Seoul National University. J.S. is also supported by a Global-LAMP Program of the NRF grant funded by the Ministry of Education (No. RS-2023-00301976). INC is supported by the National Science and Technology Council in Taiwan (Grant NSTC 114-2112-M-006-017-MY3).

\appendix

\section{Derivation of the Volume Matrix} \label{app:volume_matrix}

We derive Equation~\ref{eq:volume_matrix} in three steps. We first define the auxiliary volume $D_g$ and derive its closed form. We next construct the low-order elements of the volume matrix through explicit subtractions of the $D_g$ volumes. Finally, we generalize the subtraction to an arbitrary element $V_{i,j}$ (see also equations in the appendices of \citealp{Kriss1983} and \citealp{McLaughlin1999}). 

\begin{figure}[tp]
    \centering
    \includegraphics[width=0.8\textwidth]{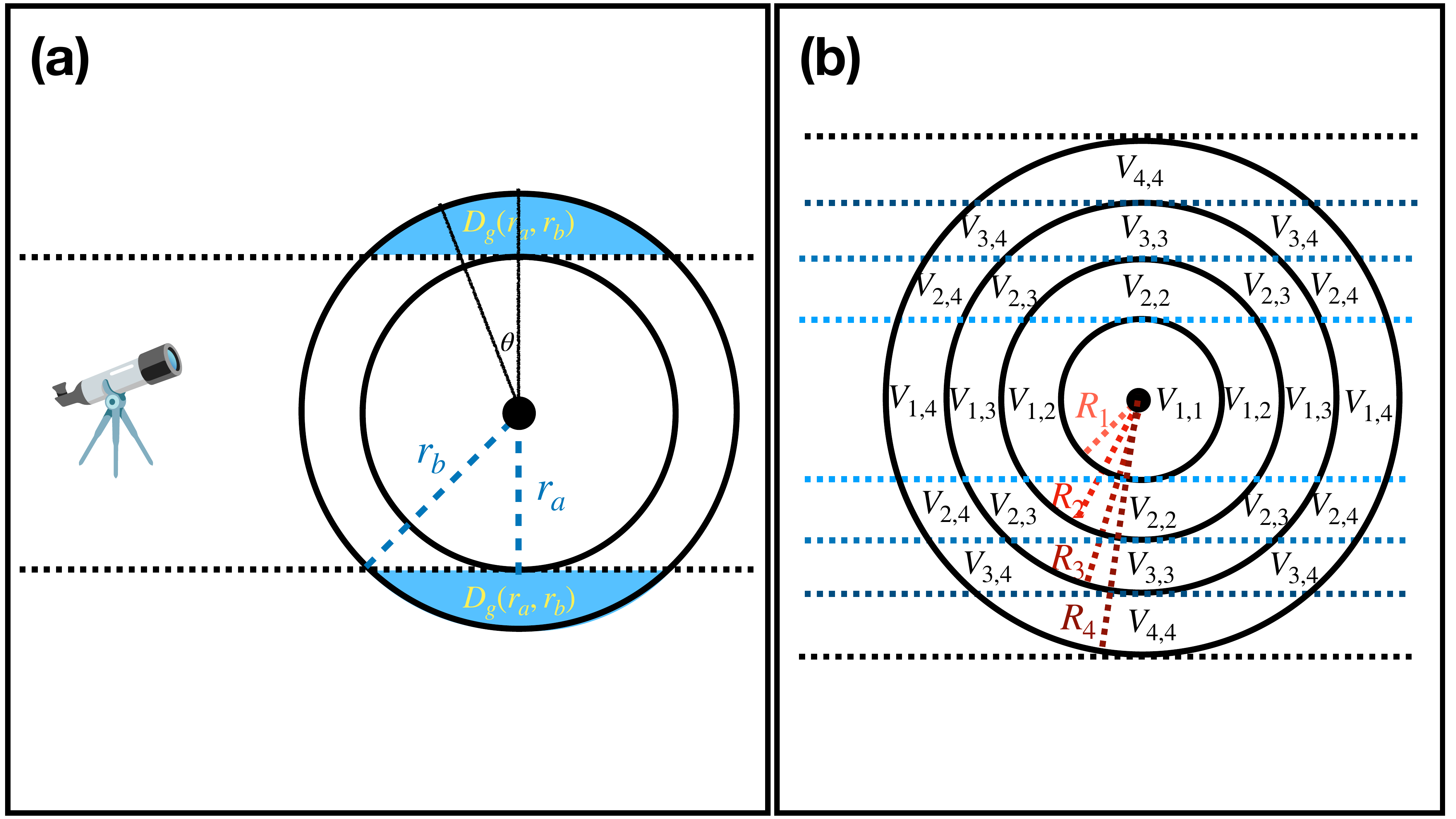}
    \caption{(a) Schematic diagram for the definition of $D_g(r_a, r_b)$. The two solid circles mark the spheres of radii $r_a$ (inner) and $r_b$ (outer), and the dotted horizontal lines mark the cylinder of radius $r_a$ along the line of sight; the telescope symbol marks the direction of the observer. $D_g(r_a, r_b)$ (blue shaded regions) is the volume inside the sphere of radius $r_b$ but outside both the sphere of radius $r_a$ and the cylinder of radius $r_a$; $\theta$ is the polar angle that spans the shaded cap. (b)~Bird's-eye (3D) view of four concentric spheres with radii $R_1 < R_2 < R_3 < R_4$; the red dashed lines mark the radii. The solid circles mark the spherical shells, and the dotted horizontal lines mark the cylinders of radii $R_1$ to $R_4$ along the line of sight. The two sets of boundaries partition the volume into the elements $V_{i,j}$, where $V_{i,j}$ is the portion of the spherical shell between $R_{j-1}$ and $R_j$ that falls inside the cylindrical annulus between $R_{i-1}$ and $R_i$. Each labeled region illustrates one element of the volume matrix in Equation~\ref{eq:volume_matrix}.}
    \label{fig:volume_matrix_diagram}
\end{figure}

We define the auxiliary volume $D_g(r_a, r_b)$; Figure~\ref{fig:volume_matrix_diagram}(a) illustrates the geometry. We consider two concentric spheres of radii $r_a$ and $r_b$ ($r_a < r_b$) that share the cluster center, and the cylinder of radius $r_a$ whose axis runs along the line of sight through the same center. $D_g(r_a, r_b)$ is the volume inside the sphere of radius $r_b$ but outside both the sphere of radius $r_a$ and the cylinder of radius $r_a$; the blue shaded regions in Figure~\ref{fig:volume_matrix_diagram}(a) mark this volume.

We compute $D_g(r_a, r_b)$ by slicing the region perpendicular to the line of sight. At height $h = r_b \sin\theta$ from the cluster center, the slice is an annulus with inner radius $r_a$ and outer radius $r_b \cos\theta$; the area of the annulus is
\begin{equation}
S(\theta) = \pi \left( r_b^2 \cos^2\theta - r_a^2 \right) .
\end{equation}
Integrating the cross-sectional area over $h$ across both caps gives
\begin{align}
D_g(r_a, r_b) &= \int S(\theta) \, dh
= 2 \int_0^{\theta_0} \pi \left( r_b^2 \cos^2\theta - r_a^2 \right) \left( r_b \cos\theta \right) d\theta , \qquad \theta_0 = \cos^{-1}\!\left( \frac{r_a}{r_b} \right) \nonumber \\
&= 2\pi \int_0^{\cos^{-1}(r_a/r_b)} \left( r_b^3 \cos^3\theta - r_a^2 r_b \cos\theta \right) d\theta \nonumber \\
&= \frac{4\pi}{3} \, r_b \sqrt{1 - \frac{r_a^2}{r_b^2}} \left( r_b^2 - r_a^2 \right) \nonumber \\
&= \frac{4}{3}\pi \left( r_b^2 - r_a^2 \right)^{\frac{3}{2}} . \label{eq:Dg_derivation}
\end{align}

Every element $V_{i,j}$ of the volume matrix in Equation~\ref{eq:volume_matrix} follows from an iterative subtraction of the $D_g$ volumes. We start with the simplest cases, the diagonal elements $V_{1,1}, V_{2,2}, V_{3,3}, \ldots$; each diagonal element equals a single $D_g$ volume. The first element $V_{1,1}$ is the volume of the sphere of radius $R_1$, which equals $D_g(0, R_1) = \frac{4}{3}\pi R_1^3$. The remaining diagonal elements likewise take the direct form $V_{i,i} = D_g(R_{i-1}, R_i)$.

The off-diagonal elements follow from subtracting known elements from a larger enclosing volume. We take $V_{1,2}$ as the first example. The sphere of radius $R_2$ contains exactly three elements: $V_{1,1}$, $V_{1,2}$, and $V_{2,2}$ (Figure~\ref{fig:volume_matrix_diagram}(b)). Subtracting the two diagonal elements from the sphere volume isolates $V_{1,2}$:
\begin{equation}
V_{1,2} = D_g(0, R_2) - V_{1,1} - V_{2,2} = D_g(0, R_2) - D_g(0, R_1) - D_g(R_1, R_2) .
\end{equation}

The element $V_{1,3}$ requires one more subtraction step. The sphere of radius $R_3$ contains six elements: $V_{1,1}$, $V_{1,2}$, $V_{1,3}$, $V_{2,2}$, $V_{2,3}$, and $V_{3,3}$. The volume $D_g(R_1, R_3)$ covers the three elements outside the cylinder of radius $R_1$, that is, $D_g(R_1, R_3) = V_{2,2} + V_{2,3} + V_{3,3}$ (Figure~\ref{fig:volume_matrix_diagram}(b)). Subtracting $D_g(R_1, R_3)$, $V_{1,2}$, and $V_{1,1}$ from the sphere volume thus isolates $V_{1,3}$:
\begin{equation}
V_{1,3} = D_g(0, R_3) - D_g(R_1, R_3) - V_{1,2} - V_{1,1} .
\end{equation}
The previous step already expresses $V_{1,2}$ through the $D_g$ volumes alone; the expression for $V_{1,3}$ thus also closes with the $D_g$ volumes. This reuse of previously derived elements is the iterative structure of Equation~\ref{eq:volume_matrix}.

The element $V_{2,3}$ follows the same logic within $D_g(R_1, R_3)$ instead of the full sphere. The volume $D_g(R_1, R_3)$ contains exactly three elements: $V_{2,2}$, $V_{2,3}$, and $V_{3,3}$. Subtracting the two diagonal elements, $V_{2,2} = D_g(R_1, R_2)$ and $V_{3,3} = D_g(R_2, R_3)$, isolates $V_{2,3}$:
\begin{equation}
V_{2,3} = D_g(R_1, R_3) - D_g(R_1, R_2) - D_g(R_2, R_3) .
\end{equation}

We now turn to a general off-diagonal element and take $V_{2,4}$ as the example. The volume $D_g(R_1, R_4)$ contains six elements: $V_{2,2}$, $V_{2,3}$, $V_{2,4}$, $V_{3,3}$, $V_{3,4}$, and $V_{4,4}$ (Figure~\ref{fig:volume_matrix_diagram}(b)). Subtracting $D_g(R_2, R_4) = V_{3,3} + V_{3,4} + V_{4,4}$ removes the elements outside the cylinder of radius $R_2$ and leaves $V_{2,2} + V_{2,3} + V_{2,4}$. The remaining two elements, $V_{2,2}$ and $V_{2,3}$, together equal $D_g(R_1, R_3) - D_g(R_2, R_3)$ by the same decomposition applied within the sphere of radius $R_3$. Thus
\begin{equation}
V_{2,4} = \left[ D_g(R_1, R_4) - D_g(R_2, R_4) \right] - \left[ D_g(R_1, R_3) - D_g(R_2, R_3) \right] ,
\end{equation}
which is Equation~\ref{eq:volume_matrix} with $i = 2$ and $j = 4$. The same subtraction generalizes to every element $V_{i,j}$ and yields the closed form of Equation~\ref{eq:volume_matrix}.

The decompositions above all follow one general relation. The volume $D_g(R_X, R_Y)$ with $R_X < R_Y$ is the sum of the volume elements inside the sphere of radius $R_Y$ but outside the cylinder of radius $R_X$:
\begin{equation}
D_g(R_X, R_Y) = \sum_{i=X+1}^{Y} \sum_{j=i}^{Y} V_{i,j} . \label{eq:Dg_sum_rule}
\end{equation}
The sum runs over the elements from $V_{X+1,X+1}$ to $V_{Y,Y}$, that is, over every element $V_{i,j}$ with $X+1 \leq i \leq j \leq Y$.

Inverting Equation~\ref{eq:Dg_sum_rule} generalizes the subtraction steps above to every volume element. The combination $D_g(R_{i-1}, R_j) - D_g(R_i, R_j)$ collects the elements of row $i$ up to column $j$, and the combination $D_g(R_{i-1}, R_{j-1}) - D_g(R_i, R_{j-1})$ collects the elements of row $i$ up to column $j-1$; the difference between the two combinations leaves the single element $V_{i,j}$:
\begin{equation}
V_{i,j} = \left[ D_g(R_{i-1}, R_{j}) - D_g(R_{i}, R_{j}) \right] - \left[ D_g(R_{i-1}, R_{j-1}) - D_g(R_{i}, R_{j-1}) \right] .
\end{equation}
Two conventions close the expression for arbitrary indices. First, $D_g(R_a, R_b) = 0$ for $R_a \geq R_b$, because the cylinder of radius $R_a$ then covers the entire sphere of radius $R_b$ and no volume remains. Second, $R_0 = 0$, so that the terms involving the innermost boundary reduce to full sphere volumes. With these conventions, every term that refers to a nonexistent region vanishes, and the expression reproduces Equation~\ref{eq:volume_matrix} for all $1 \leq i \leq j \leq N$.

\section{Analytic Integration of the Schechter Stellar Mass Function} \label{app:schechter_integration}

We estimate the stellar mass in galaxies below the MACH completeness limit by integrating the SMF analytically. We measure a cluster SMF; $\Phi(M_\star)\,dM_\star$ counts the number of member galaxies per cluster with stellar mass between $M_\star$ and $M_\star + dM_\star$. The cluster SMF follows a Schechter form \citep{Schechter1976}:
\begin{equation}
\Phi(M_\star)\, dM_\star = \frac{\phi^*}{M^*} \left( \frac{M_\star}{M^*} \right)^{\alpha} \exp\!\left( -\frac{M_\star}{M^*} \right) dM_\star ,
\label{eq:schechter}
\end{equation}
where $M^*$ is the characteristic mass, $\alpha$ is the faint-end slope, and $\phi^*$ is the normalization in units of galaxies per cluster. Because $\Phi(M_\star)\,dM_\star$ counts galaxies rather than stellar mass, we weight the SMF by $M_\star$ and evaluate two integrals: the stellar mass in galaxies below the completeness limit $M_\mathrm{cut}$, and the total stellar mass over the full mass range. The substitution $x = M_\star/M^*$, with $dM_\star = M^*\, dx$, reduces both integrals to gamma functions:
\begin{align}
\mathcal{M}(<M_\mathrm{cut}) &= \int_0^{M_\mathrm{cut}} M_\star\, \Phi(M_\star)\, dM_\star = \phi^* M^* \int_0^{x_\mathrm{cut}} x^{\alpha+1} e^{-x}\, dx = \phi^* M^*\, \gamma(\alpha+2,\, x_\mathrm{cut}) , \label{eq:schechter_partial_mass} \\
\mathcal{M}_\mathrm{tot} &= \int_0^{\infty} M_\star\, \Phi(M_\star)\, dM_\star = \phi^* M^* \int_0^{\infty} x^{\alpha+1} e^{-x}\, dx = \phi^* M^*\, \Gamma(\alpha+2) , \label{eq:schechter_total_mass}
\end{align}
where $x_\mathrm{cut} = M_\mathrm{cut}/M^*$, $\gamma(a, x)$ is the lower incomplete gamma function, and $\Gamma(a)$ is the gamma function.

The mass-weighted integral converges at the low-mass end for $\alpha > -2$. The MACH faint-end slopes give $\alpha + 2 > 0$, so $\mathcal{M}_\mathrm{tot}$ stays finite even though the integration extends to zero mass. 
The faint-end mass fraction is the ratio of Equation~\ref{eq:schechter_partial_mass} to Equation~\ref{eq:schechter_total_mass}, in which the normalization $\phi^* M^*$ cancels:
\begin{equation}
f(<M_\mathrm{cut}) = \frac{\gamma(\alpha+2,\, x_\mathrm{cut})}{\Gamma(\alpha+2)} .
\label{eq:faintend_fraction}
\end{equation}
The faint-end mass fraction thus depends only on $\alpha$ and on the ratio $M_\mathrm{cut}/M^*$, and not on the normalization $\phi^*$. \citet{Park2026b} measure the stacked MACH SMF in three projected radial bins, $R_\mathrm{proj} < 0.5\,R_{200}$, $0.5\,R_{200} \leq R_\mathrm{proj} < R_{200}$, and $R_{200} \leq R_\mathrm{proj} < 2\,R_{200}$, and fit each SMF with a single Schechter function. We adopt their incompleteness-corrected best-fit parameters: $(\alpha,\ \log M^*/\Mdot) = (-1.07,\ 10.84)$, $(-1.23,\ 10.85)$, and $(-1.19,\ 10.79)$ for the three radial bins, respectively. For the MACH completeness limit $M_\mathrm{cut} = 10^9\,\Mdot$, the three radial fits give $f = 2.0\%$ ($R_\mathrm{proj} < 0.5\,R_{200}$), $4.0\%$ ($0.5$--$1\,R_{200}$), and $3.7\%$ ($1$--$2\,R_{200}$); the fraction stays below $\sim 10\%$ for $1\sigma$ variations of the fit parameters, and the faint-end slope $\alpha$ dominates the uncertainty. The contribution of the galaxies below the completeness limit is thus limited to a few percent of the total stellar mass at all radii; we exclude these galaxies from the stellar mass profile (Section~\ref{sec:obs_results}).

\bibliographystyle{aasjournalv7}
\bibliography{ms}



\end{document}